\pdfoutput=1

\documentclass[11pt]{article}

\usepackage[final]{acl}

\usepackage{times}
\usepackage{latexsym}

\usepackage[T1]{fontenc}

\usepackage{booktabs} 
\usepackage{graphicx}
\usepackage{xcolor}
\usepackage{colortbl}
\usepackage{graphicx}
\usepackage{makecell}
\usepackage{tabularx}
\usepackage{multirow}
\usepackage{xspace}
\usepackage{enumitem}
\usepackage{float}
\usepackage{longtable}
\usepackage[most]{tcolorbox}
\usepackage{colortbl}
\usepackage{adjustbox}
\usepackage{pifont}
\usepackage{balance}
\usepackage{mfirstuc}
\usepackage{titlesec}
\usepackage{graphicx}

\usepackage{multirow}
\usepackage{graphicx}
\usepackage[normalem]{ulem}
\usepackage{tcolorbox}
\tcbuselibrary{theorems}
\usepackage{amsmath}
\usepackage{amsthm}
\usepackage{microtype}
\usepackage{amssymb}
\usepackage{times}
\usepackage{mathtools}
\usepackage{algorithm}
\usepackage{algpseudocode}
\useunder{\uline}{\ul}{}

\usepackage{color}

\newcommand{\ourtask}{interactive recommendation\xspace}

\definecolor{mycolor}{RGB}{134,150,167}
\definecolor{backred}{RGB}{255, 190, 190}
\definecolor{backblue}{RGB}{210, 230, 250}

\usepackage{color}

\newcommand{\ourmodel}{ReRec\xspace}

\definecolor{lightgreen}{RGB}{0, 135, 125}
\definecolor{lightred}{RGB}{255, 87, 51}
\title{ReRec: Reasoning-Augmented LLM-based Recommendation Assistant via Reinforcement Fine-tuning}

\author{Jiani Huang\textsuperscript{1}, Shijie Wang\textsuperscript{1}, Liangbo Ning\textsuperscript{1}, Wenqi Fan\textsuperscript{1*}, Qing Li\textsuperscript{1*} \\
\textsuperscript{1}The Hong Kong Polytechnic University \\
\texttt{jianihuang01@gmail.com; shijie.wang@connect.polyu.hk;}\\
\texttt{BigLemon1123@gmail.com;wenqifan03@gmail.com;} \\
\texttt{qing-prof.li@polyu.edu.hk} \\
}

\begin{document}
\maketitle
\renewcommand{\thefootnote}{\fnsymbol{footnote}}
\footnotetext[1]{Corresponding authors.}
\renewcommand{\thefootnote}{\arabic{footnote}}
    
\newtcbtheorem[auto counter, number within = section]{cmt}{}{                                          
      colbacktitle = mycolor, colframe = mycolor,                                                          
      colback = mycolor!10!white,                                                                   
      fonttitle=\bfseries,                                                                          
      }{t}                                                                                            

\begin{abstract}
With the rise of LLMs, there is an increasing need for intelligent recommendation assistants that can handle complex queries and provide personalized, reasoning-driven recommendations. LLM-based recommenders show potential but face challenges in multi-step reasoning, underscoring the need for reasoning-augmented systems.
To address this gap, we propose \textbf{\ourmodel}, a novel reinforcement fine-tuning (RFT) framework designed to improve LLM reasoning in complex recommendation tasks. Our framework introduces three key components: (1) \textit{Dual-Graph Enhanced Reward Shaping}, integrating recommendation metrics like NDCG@K with Query Alignment and Preference Alignment Scores to provide fine-grained reward signals for LLM optimization; (2) \textit{Reasoning-aware Advantage Estimation}, which decomposes LLM outputs into reasoning segments and penalizes incorrect steps to enhance reasoning of recommendation; and (3) \textit{Online Curriculum Scheduler}, dynamically assess query difficulty and organize training curriculum to ensure stable learning during RFT. Experiments demonstrate that \ourmodel outperforms state-of-the-art baselines and preserves core abilities like instruction-following and general knowledge. Our codes are available at \href{https://github.com/jiani-huang/ReRec}{https://github.com/jiani-huang/ReRec}.
\end{abstract}

\section{Introduction}

With the rapid advancement of AI technologies, users now expect more intelligent, context-aware recommendation systems (RecSys) that understand complex, real-time needs and provide personalized suggestions with clear reasoning~\cite{huang2025recommender, zhang2024generative}. Traditional methods, such as matrix factorization (MF) and graph neural networks (GNNs)~\cite{fan2019graph, fan2020graph}, rely on historical data like user ratings or clicks~\cite{chen2015recommender, fan2022graph}, but struggle to process natural language queries that reflect current preferences. As a result, they fail to meet the demand for intelligent recommendation assistants.

\begin{figure}
    \centering
    \includegraphics[width=1.0\linewidth]{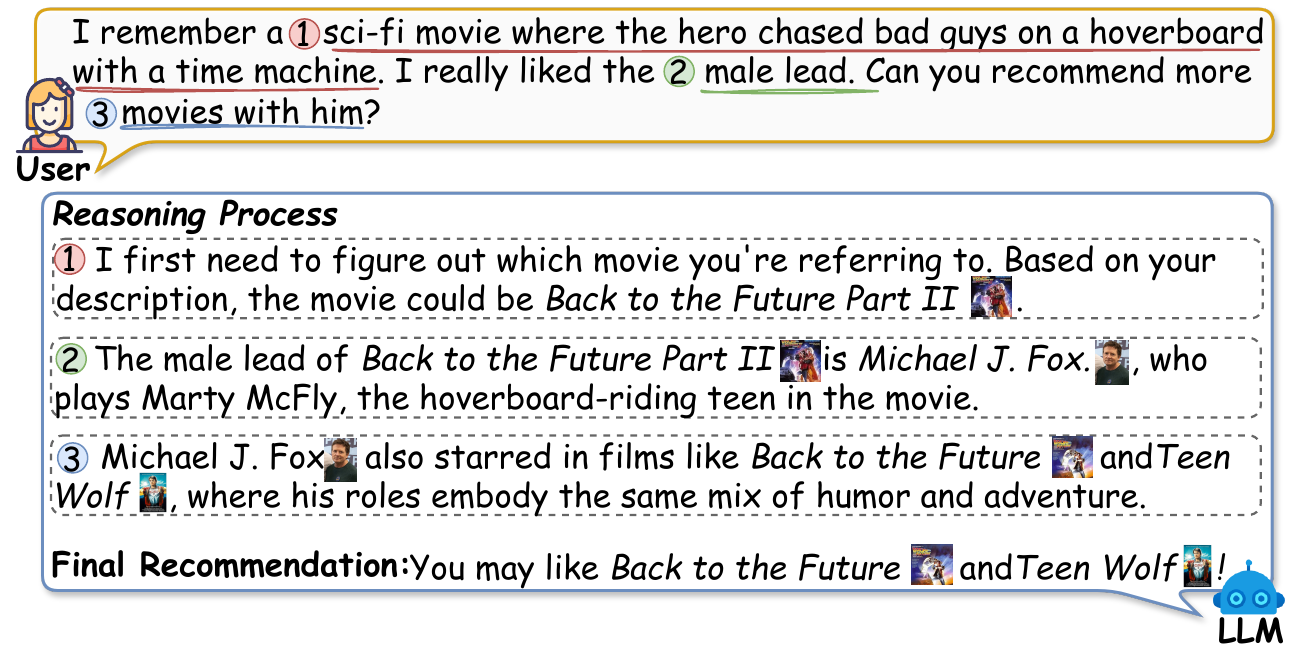}
    \caption{Example of Reasoning-Augmented LLM-based Recommendation Assistant.
    }
    \label{fig:task}
\end{figure}

The advent of large language models (LLMs) has unlocked new possibilities for intelligent, interactive recommendation assistants~\cite{zhao2024recommender,wang2025knowledge}. With their advanced language comprehension, generation abilities, and broad general knowledge~\cite{minaee2024large}, LLMs have the potential to understand natural language user queries and generate personalized recommendations. Recent work has demonstrated this potential in developing conversational recommendation systems (CRS) that engage users in multi-turn dialogues before suggesting items~\cite{yang2024behavior,liang2024llm,zhu2025llm}. However, these dialogues often involve simple and direct user queries~\cite{huang2025towards}, such as "Recommend me a sci-fi movie," which require minimal reasoning or constraints. In contrast, user queries often involve more complex queries that demand deeper reasoning for effective decision-making~\cite{ren2024explicit,wang2025adaptjobrec}. Consider the complex query illustrated in Figure~\ref{fig:task}, the assistant must first infer the correct movie from the user's description, then identify the lead actor, and finally suggest other films featuring that actor. This process requires the model to engage in multi-step reasoning, going beyond basic attribute matching or shallow semantic understanding. Existing LLM-based RecSys are limited in handling such reasoning-intensive queries due to their inadequate capacity for deep, multi-step reasoning~\cite{shi2024llm,tsai2024leveraging}. As a result, there is an urgent need for reasoning-augmented LLMs capable of addressing these complex user requests.

Recent advances in reinforcement fine-tuning (RFT) with rule-based rewards have significantly improved LLMs' reasoning and generalization for various tasks~\cite{xie2025logic,ke2025survey,zou2025traffic}. Unlike supervised fine-tuning (SFT), which requires large amounts of labeled data, RFT uses reinforcement learning (RL) to optimize the model through self-exploration. In this process, the LLM generates responses, while a reward model evaluates them, guiding the model to reinforce effective reasoning strategies. RFT offers better generalization and reduces catastrophic forgetting compared to SFT~\cite{chu2025sft}, as it focuses on active reasoning rather than memorization. 

However, despite its potential, directly applying RFT to train reasoning-augmented LLM-based recommendation assistants for complex queries presents several challenges.
One key challenge lies in developing \textbf{fine-grained reward models for complex, query-based recommendation tasks}. In general, the reward model in the RFT framework provides feedback on the recommendation quality, directly guiding the policy model updates. 
Existing studies often rely on task-specific metrics, such as NDCG, as reward signals, which can be overly stringent and sparse. 
For instance, when the LLM-based policy model generates recommendations that align with the user’s query but deviate from the ground truth, it receives the same zero reward, as responses that entirely fail to address the query. Such coarse rewards may potentially reduce the LLM-based policy model's exploration efficiency, ultimately undermining its overall performance. 
Another challenge is the \textbf{lack of supervision for the reasoning process behind the recommendations}. Recent RFT methods such as GRPO \cite{shao2024deepseekmath} typically assign a single reward score to the entire response. As all tokens share this unified score, the LLM policy cannot distinguish which specific parts of the reasoning were correct or flawed. This lack of supervision over intermediate reasoning steps makes it difficult for the model to identify and correct errors in its reasoning, such as misinterpreting user needs or overlooking key constraints of expected items. Consequently, the model struggles to improve its reasoning and may generate suboptimal recommendations \cite{yang2024behavior, zhu2025llm}.
 
To address these challenges, we propose a novel RFT-based framework (\textbf{\ourmodel}) for training a reasoning-augmented LLM-based recommendation assistant. 
In order to deliver \textbf{fine-grained reward signals}, we introduce a \textit{Dual-Graph Enhanced Reward Shaping} mechanism, which enriches traditional metrics like NDCG with two additional components: the Query Alignment Score (QAS) and the Preference Alignment Score (PAS). 
QAS evaluates how well the recommendations satisfy explicit query constraints using an item-attribute graph, while PAS assesses alignment with user preferences based on similarity to target items. 
For better \textbf{supervision of the reasoning process}, we design \textit{Reasoning-aware Advantage Estimation}, which decomposes the recommendation into reasoning steps and penalizes incorrect ones with lower advantages. 
Additionally, to mitigate the instability often associated with RL, we introduce the \textit{Online Curriculum Scheduler}, which dynamically reorders training data by prioritizing easier queries based on previous epoch performance, ensuring smoother convergence.

In summary, our contributions are:
\begin{itemize}[leftmargin=*,itemsep=0pt, parsep=0pt, topsep=0pt, partopsep=0pt]
    \item We bridge the gap between recommendation and reasoning, enabling reasoning-augmented LLM-based recommendation assistants to understand users' complex queries and provide reasonable, context-aware recommendations.
    \item We propose a reinforcement fine-tuning framework \textbf{\ourmodel} that better adapts LLMs to recommendation tasks. It aligns RL signals with recommendation goals, improves reasoning feedback, and enhances training stability, enabling more accurate and context-aware recommendations.

    \item Extensive experiments demonstrate that our method outperforms state-of-the-art baselines. Additionally, it retains strong instruction-following and reasoning capabilities, ensuring versatility for intelligent recommendations.
\end{itemize}
\section{Preliminaries}


\textbf{Problem Statement.}
Users often express preferences through complex, multifaceted natural language queries, which require the recommendation assistant to perform multi-step reasoning beyond simple keyword matching or attribute filtering to understand the user’s intent.
Formally, given a user’s query $ q $, the LLM-based recommendation assistant $ \pi_\theta $ generates a response $ o $, which includes both the reasoning process and a recommended item $ p $. The reasoning should explain why $ p $ was selected and why other items were excluded, defined as $ \pi_\theta(q) = o $.

\noindent
\textbf{Reinforcement Fine-tuning (RFT) for Recommendation.} 
To improve the reasoning capabilities of the LLM-based assistant \( \pi_\theta(q) \), reinforcement fine-tuning is typically applied to optimize its policy~\cite{wang2025re2llm,wang2025large}. Given an input query \( q \), the assistant generates multiple responses \( \{o_1, o_2, \dots, o_G\} \) based on the learned policy \( \pi_\theta \). These responses are then evaluated by a Reward Model, which assigns reward scores \( r_i \). The scores are used to compute advantages \( A_i \), indicating the quality of each response. These advantages guide the LLM policy to optimize toward improved outputs. Details of the components are provided below.
\begin{itemize}[leftmargin=*,itemsep=0.3pt, parsep=0pt, topsep=0pt, partopsep=0pt]
    \item \textit{Reward Model ($\mathcal{R}$).} For each generated output $o_i$, a predefined Reward Model $\mathcal{R}$ computes a corresponding reward value $r_i$, expressed as $ r_i=\mathcal{R}(o_i, gt) $ for each $ o_i $ with $ gt $ as the ground truth. 
    In the context of LLM-based recommendation, Normalized Discounted Cumulative Gain (NDCG)~\cite{jarvelin2002cumulated} or Recall~\cite{gunawardana2012evaluating} can be utilized as a reward model to evaluate output quality. 
    
    \item \textit{Advantage Estimation ($A$).} Research shows that directly using rewards for gradient estimation in policy optimization often results in high variance and unstable updates due to a lack of reference point~\cite{schulman2017proximal}. To address this, recent studies introduce the advantage value, which compares the actual reward to the expected reward~\cite{arulkumaran2017deep,mehta2020state}. A positive advantage encourages the policy to favor similar actions. For instance, in sampling-based advantage estimation methods~\cite{ahmadian2024back,hu2025reinforce++}, the LLM policy samples multiple responses $\{o_1, o_2, \dots, o_G\}$ for a query $q$, treating each response as a trajectory where tokens are actions. An advantage value $A_{i,t}$ is then computed for each token to identify preferred trajectories. 

    \item \textit{Training Objective ($\mathcal{J}(\theta)$).} Based on the computed advantages, the policy is optimized by maximizing following objective~\cite{shao2024deepseekmath}:
   \begin{equation}
\setlength{\jot}{1pt}
   \small
        \begin{aligned}
        & \mathcal{J}(\theta) = \mathbb{E}_{(q,gt) \sim \mathcal{D}, \{o_i\}_{i=1}^G \sim \pi_{\theta_{\text{old}}}(\cdot \mid q)} = &  \\
        & \left[ \frac{1}{N} \sum_{i=1}^G \sum_{t=1}^{|o_i|} \min \left( h_{i,t}(\theta) A_{i,t},\text{clip} \left( h_{i,t}(\theta), c_{l}, c_{h}\right) A_{i,t} \right) \right],
        \label{eq:obj}            
        \end{aligned}
    \end{equation}

    \noindent
    where \( N = \sum_{i=1}^G |o_i| \), \( h_{i,t}(\theta) = \frac{\pi_\theta(o_{i,t} \mid q, o_{i,<t})}{\pi_{\theta_{\text{old}}}(o_{i,t} \mid q, o_{i,<t})} \), the \( \text{clip} \) function represents the clipped probability ratio, and \( c_{l} = 1 - \varepsilon \), \( c_{h} = 1 + \varepsilon \). 
    
\end{itemize}



\section{Methodology}
\begin{figure*}
    \centering
    \includegraphics[width=0.9\linewidth]{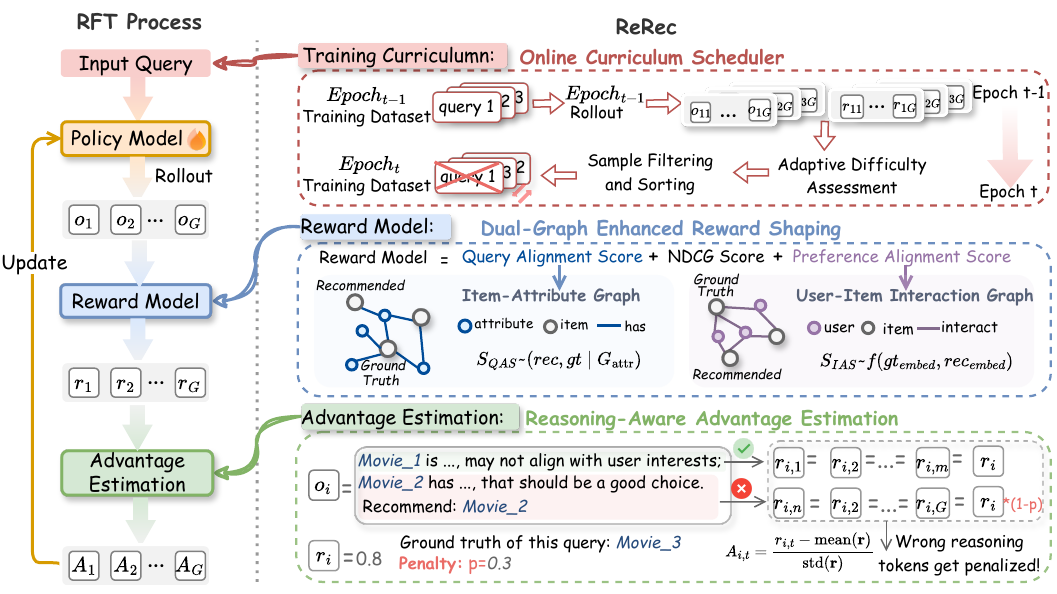}
    \caption{The overall model architecture of the proposed \ourmodel.}
    \label{fig:method}
\end{figure*}


\subsection{Overview of the Proposed \ourmodel}
\label{sec:overall}
\noindent We aims to develop a reasoning-augmented LLM-based recommendation assistant with RFT. While RFT has improved LLM reasoning in various tasks, directly applying it to query-based recommendations is challenging due to coarse reward signals and lack of supervision on intermediate reasoning. To address these issues, we propose \ourmodel, a novel RFT framework for query-based recommendation tasks. As shown in Figure~\ref{fig:method}, \ourmodel introduces a \textit{Dual-Graph Enhanced Reward Shaping} mechanism for better reward guidance, and \textit{Reasoning-aware Advantage Estimation} to supervise intermediate reasoning steps and penalize errors. An \textit{Online Curriculum Scheduler} further stabilizes training by dynamically adjusting the curriculum.

\subsection{Dual-Graph Enhanced Reward Shaping}
\label{sec:dual-graph}
\noindent 
Recent studies often use rule-based rewards in RFT to improve LLM reasoning capabilities~\cite{jin2025search,huang2025vision,wei2025webagent,luo2025gui}. For recommendation task, metrics like NDCG@K are commonly adopted as reward models. Although these metrics are established proxies for recommendation accuracy, they are unsuitable for direct use in policy optimization for query-based recommendations. Their primary limitation is evaluating only exact matches with ground-truth items, failing to assess recommendation quality comprehensively. For instance, a recommendation meeting key query constraints (e.g., genre or actor) or showing collaborative signals similar to the ground truth should be considered better than a completely unrelated recommendation, even if it does not match exactly. However, coarse-grained metrics like NDCG@K assign both the same zero reward, unable to distinguish meaningful from irrelevant recommendations. This reliance on coarse-grained rewards can destabilize RFT training, impede convergence to an optimal policy, and ultimately impair accurate reasoning and recommendations.

To overcome the above limitations, we introduce a Dual-Graph Enhanced Reward Shaping mechanism that enriches the reward space with complementary fine-grained signals. 
Specifically, in addition to the recommendation metrics $NDCG@K$, we incorporate below two auxiliary components: 

\noindent
\textbf{Query Alignment Score (QAS):}
User queries often specify constraints like genre, actor, or director that recommended items must satisfy. While the ground-truth item is a valid option, other items may also meet these criteria. Relying solely on exact matches with the ground truth for reward assignment overlooks these alternatives, penalizing reasonable recommendations and causing limited exploration of the action space and unstable training. To address this, we leverage item-attribute relationships to better evaluate whether recommended items meet query constraints. Using the ground-truth item as a reference, we assess a recommended item’s alignment by comparing shared attributes in the item-attribute graph. For an item \( p_i \) predicted by the LLM and ground truth \( gt \), given the item-attribute graph \( G_{attr} \), we compute the QAS as the proportion of shared relationships between \( p_i \) and \( gt \), as follows:
\begin{equation}
\small
    \begin{aligned}
S_{QAS}(p_i, gt) &= \frac{|R_{p_i}^{G_{attr}} \cap R_{gt}^{G_{attr}}|}{|R_{gt}^{G_{attr}}|}.
\end{aligned}
\end{equation}
\noindent
\textbf{Preference Alignment Score (PAS):} While the Query Alignment Score (QAS) evaluates whether a recommended item meets query constraints, it overlooks users’ implicit preferences beyond basic attributes. For instance, in the query “movies starring Tom Hanks,” users may prefer niche films over blockbusters. Recommending a popular film, though meeting query constraints, should incur a penalty if it misaligns with such preferences. A reward model based solely on attribute matching fails to capture these nuances. To address this, we incorporate collaborative signals from user-item interactions, reflecting co-interaction preferences~\cite{he2017neural,sarwar2001item}. We pre-train a lightweight recommender model \(\mathcal{M}\) (e.g., LightGCN~\cite{he2020lightgcn}) to generate item embeddings from the user-item interaction graph. The PAS for a recommended item \(p_i\) and ground truth \(gt\) is defined as the cosine similarity between their embeddings, as follows:
\begin{equation}
\small
\begin{aligned}
 S_{PAS}(p_i,gt) = \frac{\mathcal{M}(p_i) \cdot\mathcal{M}(g_t)}{\|\mathcal{M}(p_i)\| \|\mathcal{M}(g_t)\|}.
\end{aligned}
\end{equation}
\noindent
Formally, the final shaped reward integrates all three components and can be expressed as:
\noindent
\begin{equation}
\small
    \begin{aligned}
r_i &= \text{NDCG}+ w_1 S_{QAS} + w_2 S_{PAS},\end{aligned}
\end{equation}
where $w_1$ and $w_2$ control the influence of the auxiliary scores on the overall reward.

\subsection{Reasoning-Aware Advantage Estimation}
\label{sec:rapo}
\noindent 
Existing RFT algorithms with rule-based rewards typically assign the same advantage to all tokens based solely on the final response’s reward~\cite{shao2024deepseekmath}. This causes LLM-based recommendation systems to focus on generating the final answer, neglecting the quality of intermediate reasoning and failing to differentiate correct from incorrect steps. However, the reasoning process is essential for accurate recommendations, especially in complex scenarios that require multi-step reasoning~\cite{qu2025optimizing,wang2024openr}. Assigning the same reward to all tokens prevents the model from identifying flawed reasoning, leading to suboptimal performance. Recent research has explored process reward models to guide LLMs' intermediate reasoning~\cite{choudhury2025process,tu2025vilbench}. These methods either train dedicated models or use large pre-trained models to score reasoning steps, but both are computationally expensive and face scalability issues, limiting their practicality.

To mitigate these limitations, we propose a lightweight and effective method: \textit{Reasoning-Aware Advantage Estimation} (RAAE). RAAE provides fine-grained supervision of the reasoning process specific to recommendation tasks. Unlike conventional RFT methods, which treat all tokens equally in a reasoning trajectory, RAAE differentiates token-level contributions by penalizing tokens in reasoning steps that lead to incorrect recommendations. Specifically, we decompose the LLM's output into distinct reasoning steps and reward or penalize each segment based on its contribution to the final recommendation. 

Mathematically, given a user query \( q \) and ground truth \( gt \), the policy of LLM-based recommendation assistant generates an output \( o_i \) containing a predicted item \( p_i \) and a reasoning process. We decompose \( o_i \) into \( K \) reasoning steps via paragraph-based segmentation, formalized as:
\begin{equation}
\small
    \begin{aligned}
\mathcal{S}_i = \{s_{i,1}, \ldots, s_{i,K}\} \text{ where} \sum_{k=1}^{K} \left| s_{i,k} \right| = \left|o_i\right|,
\end{aligned}
\end{equation}
where \( s_{i,k} \) represents the \( k \)-th reasoning segment of output \( o_i \), each segment contains the reasoning step for one item. 
We assign rewards to reasoning segments based on whether they involve an incorrectly recommended item. If a segment discusses such an item that is ultimately recommended, it indicates the model failed to exclude it, and the reasoning in that segment is considered incorrect and penalized, as follows:
\begin{equation}
\small
\begin{aligned}
r_{s_{i,k}} = 
\begin{cases}
(1-w_{penalty}) \cdot r_i & \text{if } (p_i \neq gt) \land (p_i \in s_{i,k}), \\
r_i & \text{otherwise},
\end{cases}
\end{aligned}
\end{equation}
\noindent
where \( w_{penalty} \in (0,1) \) is a hyperparameter that penalizes reasoning steps with incorrect predictions, reducing rewards for associated tokens while retaining higher rewards for correct ones.
After obtaining reward of each reasoning segment, we map the segment reward $ r_{s_{i,k}} $ to each token $ t \in s_i^k $ as $ r_{i,t} $, forming $ \textbf{r} = \{r_{1,1}, \ldots, r_{G,|o_G|}\} $, and compute the advantage as $ A_{i,t} = \frac{r_{i,t} - \text{mean}(\textbf{r})}{\text{std}(\textbf{r})} $ across all tokens of outputs $ \{o_1, \ldots, o_G\} $.
The token-level advantage is then used to guide policy optimization by maximizing the objective of Equation (\ref{eq:obj}). This approach enables differential rewards across reasoning steps within the same response, thereby improving the reasoning accuracy and final recommendation performance.

\subsection{Online Curriculum Scheduler}
\label{sec:curriculum}

\noindent Training LLMs for complex recommendation tasks is challenging due to the gap between language generation and recommendation. Early in training, LLMs often struggle with complex queries, resulting in zero reward signals that hinder learning. Curriculum learning, where tasks gradually increase in difficulty, has been proposed as a solution~\cite{narvekar2020curriculum,narvekar2018learning,jiang2025self}. However, applying it to recommendation tasks is difficult, as task difficulty is harder to define compared to domains like math or code generation. Difficulty in recommendations depends on factors like the number of constraints, reasoning depth, and user expression variability~\cite{chen2025unleashing}. Additionally, traditional curriculum learning methods fail to account for the model's evolving capabilities, as they rely on static difficulty assessments made before training begins~\cite{wang2024reinforcement}.

To address these challenges, we propose an \textit{Online Curriculum Scheduler} that dynamically adjusts the training curriculum based on the policy model's evolving capabilities, which consists of three steps: 

\noindent
\textbf{Adaptive Difficulty Assessment.}
During RFT training, as the policy improves, queries that were previously difficult may become easier. It is therefore important to adaptively assess query difficulty based on the model’s evolving capabilities. This can be done by measuring the model's average performance on each query from the previous epoch. Low rewards indicate that a query is still challenging, while consistently high rewards suggest the query has been mastered and is less challenging in future iterations.
Formally, at the start of epoch \( t \), we evaluate the difficulty of samples from the previous epoch's dataset \( \mathcal{D}^{t-1} = {q_1, q_2, \dots, q_N} \), where \( q \) is a recommendation query. For each \( q\), the model generated \( G \) rollout outputs \( {o_{1}, o_{2}, \dots, o_{G}} \) in epoch \( t-1 \). The difficulty score \( d^{t-1} \) is computed as the average of the inverse rewards across all outputs:
\begin{equation}
\small
\begin{aligned}
d^{t-1} = \frac{1}{G} \sum_{i=1}^G (1 - r_{i}),
\end{aligned}
\end{equation}
\noindent
where \( r_{i} \) is the reward score for the \( i \)-th output of \( q \) in epoch \( t-1 \). 

\noindent
\textbf{Sample Filtering and Sorting.}
We apply a difficulty threshold \( \tau \) to filter out "easy" samples where \( d^{t-1} < \tau \), as consistent high performance across rollouts suggests minimal learning benefit. The remaining samples are sorted by \( d^{t-1} \) in ascending order to form the new dataset \( \mathcal{D}^t \):
\noindent
\begin{equation}
    \small
\begin{aligned}
\mathcal{D}^t = \left\{ \left( q_{(k)}, d_{(k)}^{t-1} \right) \right\}_{k=1}^m 
\text{where } \tau \leq d_{(1)}^{t-1} \leq \cdots \leq d_{(m)}^{t-1}.
\end{aligned}
\end{equation}
 This prioritizes easier samples early in epoch \( t \), fostering stable learning and gradual progression. 

\noindent
\textbf{Iterative Curriculum Update.}
The sorted \( \mathcal{D}^t \) is used for training in epoch \( t \), and the process repeats for epoch \( t+1 \) with updated difficulty \( d_n^{t} \) based on the previous rollouts. This dynamic process adapts to the model’s evolving abilities while staying efficient, as it reuses existing rollout data without relying on extra models or additional inference.

\section{Experiment}
We aim to answer the key research questions (RQs):  

\noindent\textbf{RQ1.} How does \ourmodel compare to baselines? 

\noindent\textbf{RQ2.} How effectively can \ourmodel leverage user interaction history to provide personalized recommendations?

\noindent\textbf{RQ3.} How does \ourmodel perform in generalization, e.g., cross-domain and cross-task settings?

\noindent\textbf{RQ4.} To what extent does \ourmodel retain its original knowledge and capabilities?

\subsection{Experiment Setup}
\subsubsection{Dataset.}
To evaluate our method, we conducted experiments on RecBench+~\cite{huang2025towards}, a benchmark dataset tailored for assessing complex reasoning in query-based recommendations. It covers two domains (Movie and Book) with user queries divided into five subcategories based on reasoning complexity. Details are provided in \textbf{Appendix~\ref{app:dataset}}. Data was sampled according to query category distribution, with statistics shown in Table~\ref{tab:dataset}.

\begin{table}[h]
\setlength{\intextsep}{0pt} 
  \centering
  \caption{Dataset Statistics}
    \resizebox{0.45\textwidth}{!}{
    \begin{tabular}{cccc}
    \toprule
     Category     &  Sub-category     & \multicolumn{1}{c}{Movie} & \multicolumn{1}{c}{Book} \\
    \midrule
    \multirow{3}{*}{\makecell[c]{Condition-based\\Query}} & Explicit Condition (Simple)  & 8,262      & 10,681 \\
    \cmidrule{2-4}
          & Implicit Condition (Medium)  & 5,790      & 7,741 \\
    \cmidrule{2-4}
          & Misinformed Condition (Hard)  & 5,374      & 7,890   \\
    \midrule
    \multirow{2}{*}{\makecell[c]{User Profile-based\\Query}} & Interest-based & 2,365     & 1,273 \\
    \cmidrule{2-4}
          & Demographics-based & 209     & - \\
    \midrule
    \multicolumn{2}{c}{Total} & 22,000     & 27,585\\
    \bottomrule
    \end{tabular}%
    
    }
  \label{tab:dataset}%
\end{table}

\subsubsection{Baseline Models.}
We compare our method with three categories of approaches designed to handle such queries effectively: \textbf{LLM backbones}, such as Qwen-2.5-3B-Instruct~\cite{team2024qwen2} and GPT-4o; \textbf{LLM-based CRSs}, including TallRec~\cite{bao2023tallrec}, InteRecAgent~\cite{huang2025recommender}, and CRAG~\cite{zhu2025collaborative}; and \textbf{RFT-trained Models} like GRPO~\cite{shao2024deepseekmath}, REINFORCE++~\cite{hu2025reinforce++}, and RLOO~\cite{ahmadian2024back}, where accuracy is used as the reward during training. See more information about baselines in \textbf{Appendix~\ref{app:baselines}}.

\subsubsection{Evaluation Settings.}
For each query, we generate a candidate set with 1 positive item and 19 randomly sampled negative items. Models are evaluated on Accuracy, based on selecting the correct item matching the user query.

\subsubsection{Implementations.} We select Qwen-2.5-3B-Instruct and Llama-3.2-3B-Instruct as the backbone LLM. Due to space limit, more details are provided in \textbf{Appendix~\ref{app:rq1}}.

\begin{table*}[t]
  \centering
  \caption{The overall performance of baselines and \ourmodel evaluated by Accuracy. The bold/underline values represent the best/the second-best result, respectively. 
  }
   \resizebox{0.99\textwidth}{!}{
    \begin{tabular}{c|cccccc|ccccc}
    \toprule
    \multirow{2}[2]{*}{\textbf{Category}} & \multirow{2}[2]{*}{\textbf{Model}} & \multicolumn{5}{c|}{\textbf{Movie}}             & \multicolumn{4}{c}{\textbf{Book}} \\
      &    & \multicolumn{1}{c}{Simple} & \multicolumn{1}{c}{Medium} & \multicolumn{1}{c}{Hard} & \multicolumn{1}{c}{\makecell[c]{Interest\\-based}} & \multicolumn{1}{c|}{\makecell[c]{Demographics\\-based}} & \multicolumn{1}{c}{Simple} & \multicolumn{1}{c}{Medium} & \multicolumn{1}{c}{Hard} & \multicolumn{1}{c}{\makecell[c]{Interest\\-based}} \\
    \midrule
    \multirow{5}[2]{*}{\textit{\textbf{\makecell[c]{LLM\\Backbone}}}}&   \multicolumn{1}{c}{Qwen-2.5-3B-Instruct} &  0.284     &  0.135     &  0.101     &  0.369     & 0.450      &  0.371     & 0.304      & 0.177      & 0.416 \\

  &   \multicolumn{1}{c}{Llama-3.2-3B-Instruct} &    0.107   &   0.052    & 0.029      & 0.097      &  0.193     &   0.215     & 0.138      & 0.106      & 0.254 \\

   &   \multicolumn{1}{c}{DS-R1-Distill-Qwen-7B} & 0.083      & 0.041      & 0.040      &  0.133       &  0.165     &    0.131   & 0.104     &  0.087    & 0.221   \\

   &   \multicolumn{1}{c}{GPT-4o} & 0.554 & 0.519 & 0.188 & 0.550 & 0.504 &  0.554     & 0.590      & 0.160      & 0.458\\
   &   \multicolumn{1}{c}{Deepseek-R1} & 0.537 & 0.510 & 0.200 & 0.459 & 0.425 &  0.562     &  0.530     &  0.279     & 0.505\\
    \midrule
    \multirow{3}[2]{*}{\textit{\textbf{\makecell[c]{LLM-based\\CRS}}}} &   \multicolumn{1}{c}{TallRec} &   0.537   &  0.533   &  0.284   &  0.571    &  0.509   &  0.563    &   0.591    & 0.251      & 0.477 \\
   &     \multicolumn{1}{c}{InteRecAgent} &  0.542     & 0.529      &  0.178     & 0.563      & 0.548      & 0.557      & 0.582       &  0.147     & 0.493 \\
   
  &    \multicolumn{1}{c}{CRAG} & 0.560      &  0.531    &   0.195    &  0.557     &  0.513     &  0.560     & 0.592      &  0.211     & 0.518 \\
    \midrule

   \multirow{10}[2]{*}{\textit{\textbf{\makecell[c]{RFT-trained\\Model}}}}  &  \multicolumn{10}{c}{\cellcolor{gray!20}{Qwen-2.5-3B-Instruct}} \\
   &   \multicolumn{1}{c}{GRPO} & 0.549  & 0.502 & 0.461 & \underline{0.579} & \underline{0.648} & \underline{0.563} & 0.630 & \underline{0.552} & 0.699 \\
    
    &  \multicolumn{1}{c}{REINFORCE++} & \underline{0.578} & 0.523  & 0.506 & 0.556 & 0.637 & 0.553   & 0.618 & 0.510 & \underline{0.716} \\

    &  \multicolumn{1}{c}{RLOO} & 0.560  & 0.495  & \underline{0.529} & 0.573 & 0.614  & \textbf{0.567}   & \underline{0.649} & 0.532  & \underline{0.716}\\
    
   &   \multicolumn{1}{c}{\textbf{\ourmodel}} & \textbf{0.595} & \textbf{0.548} & \textbf{0.547}  & \textbf{0.588} & \textbf{0.670}  & 0.555 & \textbf{0.655} & \textbf{0.562} & \textbf{0.746} \\

     &  \multicolumn{10}{c}{\cellcolor{gray!20}{Llama-3.2-3B-Instruct}} \\
    &  \multicolumn{1}{c}{GRPO} &  0.686    &  0.600     &   \underline{0.644}    & 0.651      &   0.642    &  \underline{0.664} & 0.757      &  \underline{0.713}     & 0.786 \\
    
    &  \multicolumn{1}{c}{REINFORCE++} &  \underline{0.699}     &  \underline{0.623}     &    0.597   & \underline{0.676}      &    \underline{0.771}  &  0.661      &  0.768     &  0.697     & \underline{0.795} \\

   &   \multicolumn{1}{c}{RLOO}&  0.693     &  0.609     & 0.614    &   0.627    &  0.688     &    0.660  &   \underline{0.774 }   &   0.704    &  0.794\\
    
   &   \multicolumn{1}{c}{\textbf{\ourmodel}} &  \textbf{0.748}     &  \textbf{0.700}     & \textbf{ 0.729 }    &  \textbf{ 0.719}    &   \textbf{0.800 }   &    \textbf{0.671}  &   \textbf{0.782}    &   \textbf{0.750}    & \textbf{0.811} \\
    
    \bottomrule
    \end{tabular}%
    
    }
  \label{tab:rq1}%
\end{table*}%
\subsection{Overall Performance (RQ1)} 
\noindent 
We compare our method with various LLM backbones, LLM-based CRS models, and RFT-trained models using the same test set. The results are presented in Table~\ref{tab:rq1}. From the table, it is evident that LLM backbones, such as Qwen-2.5-3B-Instruct and Llama-3.2-3B-Instruct, perform significantly worse than more advanced closed-source models like GPT-4o and DeepSeek-R1. While these LLM backbones excel on simpler tasks, such as Explicit Condition-based Query, their performance significantly declines on more complex tasks requiring deeper reasoning, like Misinformed Condition-based (Hard) Query. LLM-based CRS models achieve slightly better performance across all query types compared to closed-source LLMs. RFT-trained models show the most significant improvement, with \ourmodel outperforming other RFT-based models across domains and query types.
For example, on the Llama-3.2-3B-Instruct model, our \ourmodel achieves a performance improvement of 3.76\% to 13.2\% over the second-best method in the Movie domain.
In particular, on more challenging tasks like Misinformed-based (Hard) Query, \ourmodel demonstrates a remarkable 440\% improvement over the untrained model. 
This significant gain highlights a marked improvement in LLM's reasoning abilities, enabling it to identify misleading information within the query and better understand the user's true intentions. 

\subsection{Personalized Recommendation (RQ2)}
\noindent 
In real-world settings, multiple items may satisfy a given query. To assess the model’s ability to personalize recommendations based on user history, we conduct experiments in a personalized scenario. Instead of randomly sampling 19 negative items, we select K = 3 “hard negatives” that meet the query criteria but are not the ground truth. We then compare model performance in two settings: (1) using only the query, and (2) using both the query and the user’s interaction history. A model capable of leveraging historical preferences should distinguish the positive item from the hard negatives.
\begin{figure}[b]
    \centering
  \includegraphics[width=1\linewidth]{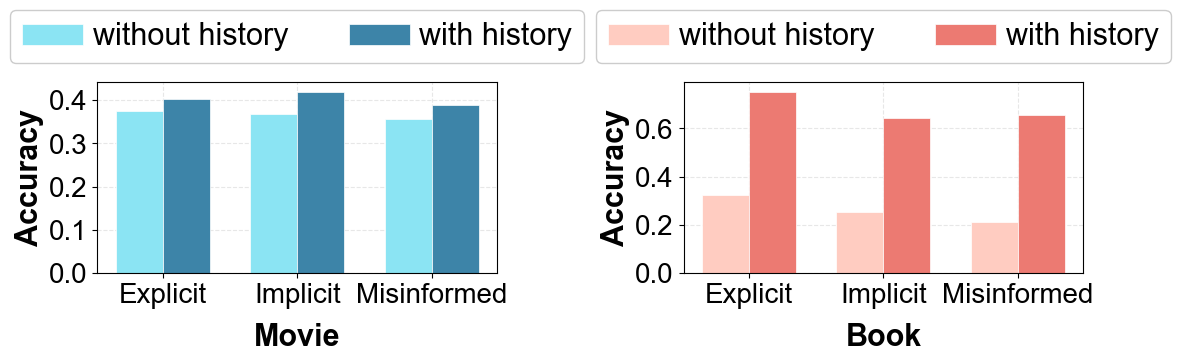}
    \caption{Performance on personalized recommendation}
    \label{fig:rq2}
\end{figure}

As shown in Figure~\ref{fig:rq2}, providing user interaction history improves model performance across different categories and domains. This suggests that the model considers both the query and the user’s preferences when selecting items, enabling it to better exclude hard negatives. These results show that \ourmodel effectively uses user history to reason preferences and generate more personalized recommendations.



\begin{table}[t]
  \centering
  \caption{Performance in the generalization (cross-domain) setting. \textbf{Bold:} Accuracy on new domain.
}
  \resizebox{0.45\textwidth}{!}{
  \small
    \begin{tabular}{ccccc}
    \toprule
    \multirow{2}[2]{*}{\makecell[c]{Training\\Dataset}} & \multicolumn{1}{c}{\multirow{2}[2]{*}{Method}} & \multicolumn{1}{c}{\multirow{2}[2]{*}{Base Model}} & \multicolumn{2}{c}{Testing Dataset} \\ \cline{4-5}
          &       &       & \multicolumn{1}{c}{Movie} & \multicolumn{1}{c}{Book} \\
    \midrule
    \multirow{4}[2]{*}{zero-shot} & prompt & Qwen  &  0.240   &  0.301 \\
          & prompt & Llama &  0.078    & 0.168  \\
          & prompt & GPT-4o &  0.470    & 0.453  \\
          & prompt & DeepSeek-R1 &  0.441   &  0.474 \\
    \midrule
    \multirow{2}[2]{*}{Movie} & \ourmodel & Qwen  & 0.567  & \textbf{0.434}  \\
          & \ourmodel & Llama & 0.727     & \textbf{0.494}  \\
    \midrule
    \multirow{2}[2]{*}{Book} & \ourmodel & Qwen  & \textbf{0.406} & 0.598  \\
          & \ourmodel & Llama &   \textbf{ 0.448 }  &   0.733 \\
    \bottomrule
    \end{tabular}%
    }
  \label{tab:rq3}%
\end{table}%
\subsection{Generalization Capability (RQ3)}
\noindent 
The generalization capability is crucial for LLM-based recommendation assistants, as users may request recommendations across a wide range of domains or require the model to perform different types of recommendation tasks. To effectively handle this problem, the assistant needs to generalize well, adapting to new contexts and tasks while maintaining high performance. To explore the generalization capabilities of \ourmodel, we conducted experiments to assess its cross-domain and cross-task generalization.
For \textbf{cross-domain generalization}, we evaluated the model’s ability to transfer knowledge between domains. For example, a model trained on the Movie domain was applied to zero-shot recommendations on Book data. As shown in Table~\ref{tab:rq3}, \ourmodel (with Llama backbone) achieves a score of 0.494 when trained on Movie and tested on Book, a 181\% relative improvement over the base model (0.168), outperforming strong models like GPT-4o and DeepSeek-R1. A similar trend was observed when transferring from Book to Movie. These results demonstrate that \ourmodel effectively generalizes across domains, capturing reasoning patterns that are not domain-specific.


\begin{table}[t]
  \centering
  \caption{Performance on transferring to sequential recommendation.}
    \small
     \resizebox{0.49\textwidth}{!}{
    \begin{tabular}{cc}
    \toprule
    Model & Accuracy \\
    \midrule
    Llama-3.2-3B-Instruct & 0.120 \\
    Qwen-2.5-3B-Instruct & 0.286 \\
    GRU4Rec & 0.658 \\
    SASRec & \textbf{0.673} (\textit{best}) \\
    \midrule
    ReRec-Qwen & 0.591 (vs. Qwen +107\% | vs. \textit{best} 87.8\%) \\
    ReRec-Llama & 0.595 (vs. Llama +396\% | vs. \textit{best} 88.4\%) \\
    \bottomrule
    \end{tabular}%
    }
  \label{tab:rq3-2}%
\end{table}%

For \textbf{cross-task generalization}, we evaluated the model on a different task: sequential recommendation, where it predicts the next item based on a user’s interaction history. 
More details about experiments are provided in \textbf{Appendix~\ref{app:rq3}}. Table~\ref{tab:rq3-2} indicates that \ourmodel, trained on our complex query-based task, transfers well to sequential recommendation, achieving substantial gains over base models (Qwen +107\%, Llama +396\%). Moreover, \ourmodel achieves up to 90\% of the performance of specialized models like SASRec.


\subsection{Capability Retention (RQ4)}
\noindent 
Beyond recommendation accuracy, a reasoning-augmented LLM-based assistant also needs world knowledge and instruction-following skills to interpret user queries and provide relevant, personalized responses. Preserving the base model’s original capabilities after RFT is therefore critical. Prior work has shown that SFT can cause catastrophic forgetting, where task-specific learning leads to loss of previously acquired knowledge~\cite{chu2025sft}. We evaluate how our method, compared to SFT, maintains the model’s original abilities across four key dimensions relevant to \ourtask: \textbf{Reasoning}, \textbf{Instruction-Following}, \textbf{Multiple Choice Question (MCQ) answering}, and \textbf{Knowledge}. Details on the SFT procedure and evaluation benchmarks are given in \textbf{Appendix~\ref{app:rq4}}.

\begin{figure}[t]
    \centering
    \includegraphics[width=1\linewidth]{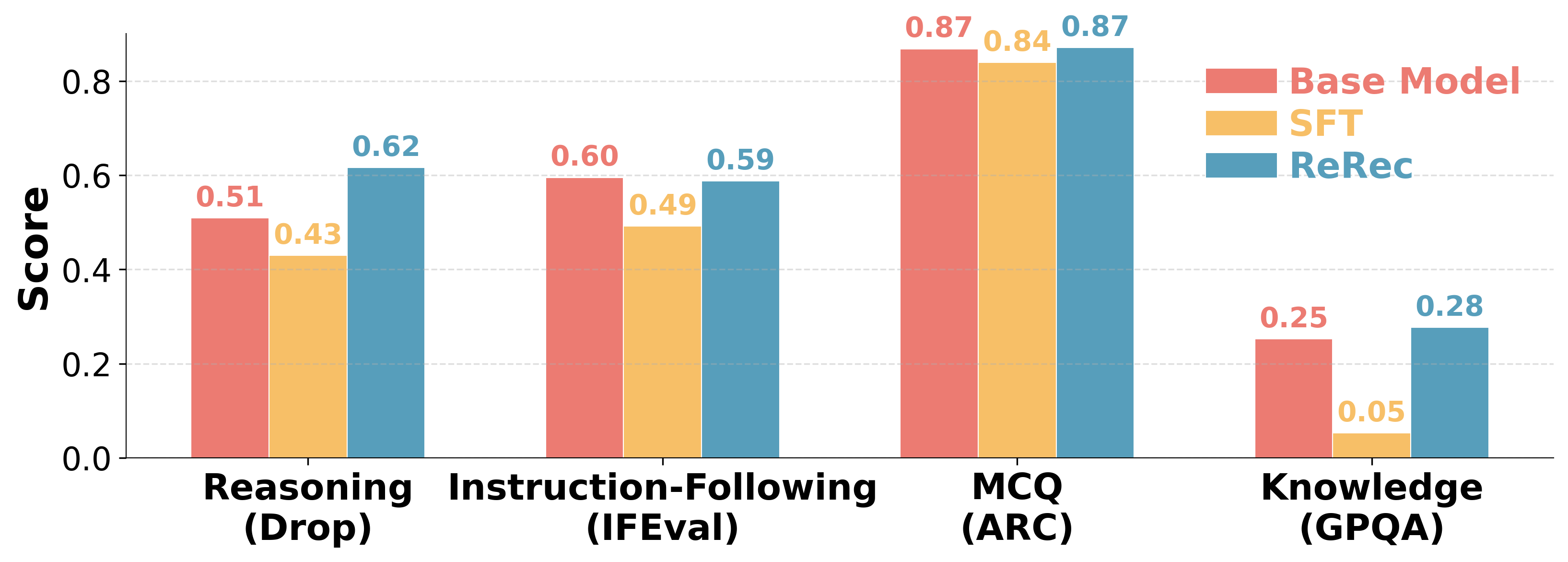}
    \caption{Knowledge and Capability Retention}
    \label{fig:rq4}
\end{figure}

As shown in Figure~\ref{fig:rq4}, our model preserves the original model’s capabilities across all four dimensions with minimal loss. Remarkably, its reasoning ability improved by 21.6\% over the base model. In contrast, the SFT-trained model suffered substantial declines, especially in Reasoning and Knowledge, which dropped by 15.7\% and 80\%, respectively. Such degradation can harm LLM-based recommendation assistants, making them appear less intelligent and reducing user satisfaction and trust.

\subsection{Ablation Study}
\begin{figure}[b]
    \centering
    \includegraphics[width=1.0\linewidth]{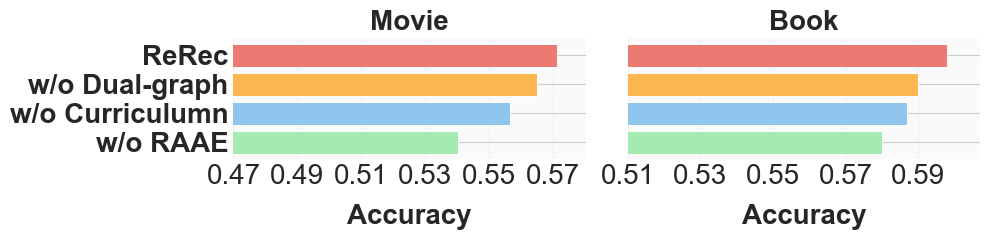}
    \caption{Ablation of \ourmodel}
    \label{fig:ablation}
\end{figure}

\noindent 
To evaluate each module’s impact, we performed an ablation study by removing components one at a time. As shown in Figure~\ref{fig:ablation}, the full model \ourmodel achieves the highest accuracy. Removing each component reduces performance, while the largest drop occurs when RAEE is removed. 

\section{Related Work}
\textbf{LLM-based Recommendation.}

\noindent
With the rapid advancement of LLMs, researchers have increasingly explored their applications in recommender systems~\cite{zhao2024recommender}. For example, ~\citet{lyu2023llm} used LLMs to generate detailed, context-aware user and item profiles from historical interactions, enhancing the expressiveness of input features. TallRec~\cite{bao2023tallrec} applied instruction fine-tuning to LLMs for recommendation tasks.
LLMs have also been studied for conversational recommendation systems (CRS). ~\citet{friedman2023leveraging} proposed RecLLM, a LaMDA-based system for YouTube video recommendation that captures user preferences, manages dialogue flexibly, and produces explainable recommendations. Similarly, Feng et al.~\cite{feng2023large} developed a hybrid architecture combining LLMs with domain-specific expert models to improve CRS performance.
Despite these advances, most existing studies focus on short, template-based dialogues with simple intents (e.g., “I want a [genre] movie”), which limits query diversity and ignores the complex reasoning scenarios. To address this, we use RecBench+~\cite{huang2025towards}, a recent benchmark featuring user queries of varying reasoning difficulty to more effectively evaluate recommendation reasoning.

\noindent \textbf{Reinforcement Learning for LLM Reasoning.} 
Recent studies show that reinforcement fine-tuning can greatly enhance the reasoning ability of LLMs. Models such as Deepseek-R1~\cite{guo2025deepseek} and Kimi K1.5~\cite{team2025kimi} use RL algorithms like GRPO~\cite{shao2024deepseekmath} to enable multi-step reasoning. Several GRPO variants, including DAPO~\cite{yu2025dapo} and Dr.GRPO~\cite{shao2024deepseekmath}, have further improved the efficiency and effectiveness of RL in post-training. These RL-based methods have been successfully applied in video understanding~\cite{feng2025video,jiang2024prior}, audio processing~\cite{rouditchenko2025omni}, and robotics~\cite{kim2025robot}, significantly enhancing reasoning and generalization. However, in recommendation domain, RL research has mainly focused on sequential recommendation, with limited attention to reasoning-intensive query-based recommendations.

\section{Conclusion}
In this paper, we present \textbf{\ourmodel}, a reinforcement fine-tuning framework that enhances LLM-based recommendation assistants with improved reasoning. To better adapt RFT to recommendation tasks, we introduce fine-grained reward shaping and reasoning-aware advantage estimation. Extensive experiments demonstrate that \textbf{\ourmodel} outperforms state-of-the-art baselines in recommendation accuracy, while preserving instruction-following and general knowledge capabilities.


\section{Limitations}
While our \ourmodel framework effectively enhances reasoning in single-turn query-based recommendations, it does not account for multi-turn dialogues, which are common in real-world conversational recommendation systems. This may limit its applicability in scenarios requiring ongoing user interactions and context accumulation. Future work could extend the framework to incorporate multi-turn capabilities, such as maintaining conversation history and adapting rewards dynamically across interactions.

\bibliography{sample-base}

\begin{thebibliography}{63}
\providecommand{\natexlab}[1]{#1}

\bibitem[{Ahmadian et~al.(2024)Ahmadian, Cremer, Gall{\'e}, Fadaee, Kreutzer, Pietquin, {\"U}st{\"u}n, and Hooker}]{ahmadian2024back}
Arash Ahmadian, Chris Cremer, Matthias Gall{\'e}, Marzieh Fadaee, Julia Kreutzer, Olivier Pietquin, Ahmet {\"U}st{\"u}n, and Sara Hooker. 2024.
\newblock Back to basics: Revisiting reinforce style optimization for learning from human feedback in llms.
\newblock \emph{arXiv preprint arXiv:2402.14740}.

\bibitem[{Arulkumaran et~al.(2017)Arulkumaran, Deisenroth, Brundage, and Bharath}]{arulkumaran2017deep}
Kai Arulkumaran, Marc~Peter Deisenroth, Miles Brundage, and Anil~Anthony Bharath. 2017.
\newblock Deep reinforcement learning: A brief survey.
\newblock \emph{IEEE signal processing magazine}, 34(6):26--38.

\bibitem[{Bao et~al.(2023)Bao, Zhang, Zhang, Wang, Feng, and He}]{bao2023tallrec}
Keqin Bao, Jizhi Zhang, Yang Zhang, Wenjie Wang, Fuli Feng, and Xiangnan He. 2023.
\newblock Tallrec: An effective and efficient tuning framework to align large language model with recommendation.
\newblock In \emph{Proceedings of the 17th ACM conference on recommender systems}, pages 1007--1014.

\bibitem[{Chen et~al.(2025)Chen, Zhang, Langren{\'e}, and Zhu}]{chen2025unleashing}
Banghao Chen, Zhaofeng Zhang, Nicolas Langren{\'e}, and Shengxin Zhu. 2025.
\newblock Unleashing the potential of prompt engineering for large language models.
\newblock \emph{Patterns}.

\bibitem[{Chen et~al.(2015)Chen, Chen, and Wang}]{chen2015recommender}
Li~Chen, Guanliang Chen, and Feng Wang. 2015.
\newblock Recommender systems based on user reviews: the state of the art.
\newblock \emph{User Modeling and User-Adapted Interaction}, 25(2):99--154.

\bibitem[{Choudhury(2025)}]{choudhury2025process}
Sanjiban Choudhury. 2025.
\newblock Process reward models for llm agents: Practical framework and directions.
\newblock \emph{arXiv preprint arXiv:2502.10325}.

\bibitem[{Chu et~al.(2025)Chu, Zhai, Yang, Tong, Xie, Schuurmans, Le, Levine, and Ma}]{chu2025sft}
Tianzhe Chu, Yuexiang Zhai, Jihan Yang, Shengbang Tong, Saining Xie, Dale Schuurmans, Quoc~V Le, Sergey Levine, and Yi~Ma. 2025.
\newblock Sft memorizes, rl generalizes: A comparative study of foundation model post-training.
\newblock \emph{arXiv preprint arXiv:2501.17161}.

\bibitem[{Fan et~al.(2022)Fan, Liu, Jin, Zhao, Tang, and Li}]{fan2022graph}
Wenqi Fan, Xiaorui Liu, Wei Jin, Xiangyu Zhao, Jiliang Tang, and Qing Li. 2022.
\newblock Graph trend filtering networks for recommendation.
\newblock In \emph{Proceedings of the 45th International ACM SIGIR Conference on Research and Development in Information Retrieval}, pages 112--121.

\bibitem[{Fan et~al.(2019)Fan, Ma, Li, He, Zhao, Tang, and Yin}]{fan2019graph}
Wenqi Fan, Yao Ma, Qing Li, Yuan He, Eric Zhao, Jiliang Tang, and Dawei Yin. 2019.
\newblock Graph neural networks for social recommendation.
\newblock In \emph{The world wide web conference}, pages 417--426.

\bibitem[{Fan et~al.(2020)Fan, Ma, Li, Wang, Cai, Tang, and Yin}]{fan2020graph}
Wenqi Fan, Yao Ma, Qing Li, Jianping Wang, Guoyong Cai, Jiliang Tang, and Dawei Yin. 2020.
\newblock A graph neural network framework for social recommendations.
\newblock \emph{IEEE Transactions on Knowledge and Data Engineering}.

\bibitem[{Feng et~al.(2025)Feng, Gong, Li, Guo, Wang, Peng, Wu, Zhang, Wang, and Yue}]{feng2025video}
Kaituo Feng, Kaixiong Gong, Bohao Li, Zonghao Guo, Yibing Wang, Tianshuo Peng, Junfei Wu, Xiaoying Zhang, Benyou Wang, and Xiangyu Yue. 2025.
\newblock Video-r1: Reinforcing video reasoning in mllms.
\newblock \emph{arXiv preprint arXiv:2503.21776}.

\bibitem[{Feng et~al.(2023)Feng, Liu, Xue, Cai, Hu, Jiang, Gai, and Sun}]{feng2023large}
Yue Feng, Shuchang Liu, Zhenghai Xue, Qingpeng Cai, Lantao Hu, Peng Jiang, Kun Gai, and Fei Sun. 2023.
\newblock A large language model enhanced conversational recommender system.
\newblock \emph{arXiv preprint arXiv:2308.06212}.

\bibitem[{Friedman et~al.(2023)Friedman, Ahuja, Allen, Tan, Sidahmed, Long, Xie, Schubiner, Patel, Lara et~al.}]{friedman2023leveraging}
Luke Friedman, Sameer Ahuja, David Allen, Zhenning Tan, Hakim Sidahmed, Changbo Long, Jun Xie, Gabriel Schubiner, Ajay Patel, Harsh Lara, and 1 others. 2023.
\newblock Leveraging large language models in conversational recommender systems.
\newblock \emph{arXiv preprint arXiv:2305.07961}.

\bibitem[{Gunawardana et~al.(2012)Gunawardana, Shani, and Yogev}]{gunawardana2012evaluating}
Asela Gunawardana, Guy Shani, and Sivan Yogev. 2012.
\newblock Evaluating recommender systems.
\newblock In \emph{Recommender systems handbook}, pages 547--601. Springer.

\bibitem[{Guo et~al.(2025)Guo, Yang, Zhang, Song, Zhang, Xu, Zhu, Ma, Wang, Bi et~al.}]{guo2025deepseek}
Daya Guo, Dejian Yang, Haowei Zhang, Junxiao Song, Ruoyu Zhang, Runxin Xu, Qihao Zhu, Shirong Ma, Peiyi Wang, Xiao Bi, and 1 others. 2025.
\newblock Deepseek-r1: Incentivizing reasoning capability in llms via reinforcement learning.
\newblock \emph{arXiv preprint arXiv:2501.12948}.

\bibitem[{He et~al.(2020)He, Deng, Wang, Li, Zhang, and Wang}]{he2020lightgcn}
Xiangnan He, Kuan Deng, Xiang Wang, Yan Li, Yongdong Zhang, and Meng Wang. 2020.
\newblock Lightgcn: Simplifying and powering graph convolution network for recommendation.
\newblock In \emph{Proceedings of the 43rd International ACM SIGIR conference on research and development in Information Retrieval}, pages 639--648.

\bibitem[{He et~al.(2017)He, Liao, Zhang, Nie, Hu, and Chua}]{he2017neural}
Xiangnan He, Lizi Liao, Hanwang Zhang, Liqiang Nie, Xia Hu, and Tat-Seng Chua. 2017.
\newblock Neural collaborative filtering.
\newblock In \emph{Proceedings of the 26th international conference on world wide web}, pages 173--182.

\bibitem[{Hidasi et~al.(2015)Hidasi, Karatzoglou, Baltrunas, and Tikk}]{hidasi2015session}
Bal{\'a}zs Hidasi, Alexandros Karatzoglou, Linas Baltrunas, and Domonkos Tikk. 2015.
\newblock Session-based recommendations with recurrent neural networks.
\newblock \emph{arXiv preprint arXiv:1511.06939}.

\bibitem[{Hu(2025)}]{hu2025reinforce++}
Jian Hu. 2025.
\newblock Reinforce++: A simple and efficient approach for aligning large language models.
\newblock \emph{arXiv preprint arXiv:2501.03262}.

\bibitem[{Huang et~al.(2025{\natexlab{a}})Huang, Wang, Ning, Fan, Wang, Yin, and Li}]{huang2025towards}
Jiani Huang, Shijie Wang, Liang-bo Ning, Wenqi Fan, Shuaiqiang Wang, Dawei Yin, and Qing Li. 2025{\natexlab{a}}.
\newblock Towards next-generation recommender systems: A benchmark for personalized recommendation assistant with llms.
\newblock \emph{arXiv preprint arXiv:2503.09382}.

\bibitem[{Huang et~al.(2025{\natexlab{b}})Huang, Jia, Zhai, Cao, Ye, Zhao, Xu, Hu, and Lin}]{huang2025vision}
Wenxuan Huang, Bohan Jia, Zijie Zhai, Shaosheng Cao, Zheyu Ye, Fei Zhao, Zhe Xu, Yao Hu, and Shaohui Lin. 2025{\natexlab{b}}.
\newblock Vision-r1: Incentivizing reasoning capability in multimodal large language models.
\newblock \emph{arXiv preprint arXiv:2503.06749}.

\bibitem[{Huang et~al.(2025{\natexlab{c}})Huang, Lian, Lei, Yao, Lian, and Xie}]{huang2025recommender}
Xu~Huang, Jianxun Lian, Yuxuan Lei, Jing Yao, Defu Lian, and Xing Xie. 2025{\natexlab{c}}.
\newblock Recommender ai agent: Integrating large language models for interactive recommendations.
\newblock \emph{ACM Transactions on Information Systems}, 43(4):1--33.

\bibitem[{J{\"a}rvelin and Kek{\"a}l{\"a}inen(2002)}]{jarvelin2002cumulated}
Kalervo J{\"a}rvelin and Jaana Kek{\"a}l{\"a}inen. 2002.
\newblock Cumulated gain-based evaluation of ir techniques.
\newblock \emph{ACM Transactions on Information Systems (TOIS)}, 20(4):422--446.

\bibitem[{Jiang et~al.(2025)Jiang, Qian, Wu, Huang, Li, Wu, and Wei}]{jiang2025self}
Yiyang Jiang, Guangwu Qian, Jiaxin Wu, Qi~Huang, Qing Li, Yongkang Wu, and Xiao-Yong Wei. 2025.
\newblock Self-paced learning for images of antinuclear antibodies.
\newblock \emph{IEEE Transactions on Medical Imaging}.

\bibitem[{Jiang et~al.(2024)Jiang, Zhang, Zhang, Wei, Chen, and Li}]{jiang2024prior}
Yiyang Jiang, Wengyu Zhang, Xulu Zhang, Xiao-Yong Wei, Chang~Wen Chen, and Qing Li. 2024.
\newblock Prior knowledge integration via llm encoding and pseudo event regulation for video moment retrieval.
\newblock In \emph{Proceedings of the 32nd ACM International Conference on Multimedia}, pages 7249--7258.

\bibitem[{Jin et~al.(2025)Jin, Zeng, Yue, Yoon, Arik, Wang, Zamani, and Han}]{jin2025search}
Bowen Jin, Hansi Zeng, Zhenrui Yue, Jinsung Yoon, Sercan Arik, Dong Wang, Hamed Zamani, and Jiawei Han. 2025.
\newblock Search-r1: Training llms to reason and leverage search engines with reinforcement learning.
\newblock \emph{arXiv preprint arXiv:2503.09516}.

\bibitem[{Kang and McAuley(2018)}]{kang2018self}
Wang-Cheng Kang and Julian McAuley. 2018.
\newblock Self-attentive sequential recommendation.
\newblock In \emph{2018 IEEE international conference on data mining (ICDM)}, pages 197--206. IEEE.

\bibitem[{Ke et~al.(2025)Ke, Jiao, Ming, Nguyen, Xu, Long, Li, Qin, Wang, Savarese et~al.}]{ke2025survey}
Zixuan Ke, Fangkai Jiao, Yifei Ming, Xuan-Phi Nguyen, Austin Xu, Do~Xuan Long, Minzhi Li, Chengwei Qin, Peifeng Wang, Silvio Savarese, and 1 others. 2025.
\newblock A survey of frontiers in llm reasoning: Inference scaling, learning to reason, and agentic systems.
\newblock \emph{arXiv preprint arXiv:2504.09037}.

\bibitem[{Kim et~al.(2025)Kim, Park, Jang, Shin, Kim, and Seo}]{kim2025robot}
Dongyoung Kim, Sumin Park, Huiwon Jang, Jinwoo Shin, Jaehyung Kim, and Younggyo Seo. 2025.
\newblock Robot-r1: Reinforcement learning for enhanced embodied reasoning in robotics.
\newblock \emph{arXiv preprint arXiv:2506.00070}.

\bibitem[{Liang et~al.(2024)Liang, Jin, Wang, Fan, Xia, Chen, and Yin}]{liang2024llm}
Tingting Liang, Chenxin Jin, Lingzhi Wang, Wenqi Fan, Congying Xia, Kai Chen, and Yuyu Yin. 2024.
\newblock Llm-redial: a large-scale dataset for conversational recommender systems created from user behaviors with llms.
\newblock In \emph{Findings of the Association for Computational Linguistics ACL 2024}, pages 8926--8939.

\bibitem[{Luo et~al.(2025)Luo, Wang, He, and Xia}]{luo2025gui}
Run Luo, Lu~Wang, Wanwei He, and Xiaobo Xia. 2025.
\newblock Gui-r1: A generalist r1-style vision-language action model for gui agents.
\newblock \emph{arXiv preprint arXiv:2504.10458}.

\bibitem[{Lyu et~al.(2023)Lyu, Jiang, Zeng, Xia, Wang, Zhang, Chen, Leung, Tang, and Luo}]{lyu2023llm}
Hanjia Lyu, Song Jiang, Hanqing Zeng, Yinglong Xia, Qifan Wang, Si~Zhang, Ren Chen, Christopher Leung, Jiajie Tang, and Jiebo Luo. 2023.
\newblock Llm-rec: Personalized recommendation via prompting large language models.
\newblock \emph{arXiv preprint arXiv:2307.15780}.

\bibitem[{Mehta(2020)}]{mehta2020state}
Deepanshu Mehta. 2020.
\newblock State-of-the-art reinforcement learning algorithms.
\newblock \emph{International Journal of Engineering Research and Technology}, 8(1):717--722.

\bibitem[{Minaee et~al.(2024)Minaee, Mikolov, Nikzad, Chenaghlu, Socher, Amatriain, and Gao}]{minaee2024large}
Shervin Minaee, Tomas Mikolov, Narjes Nikzad, Meysam Chenaghlu, Richard Socher, Xavier Amatriain, and Jianfeng Gao. 2024.
\newblock Large language models: A survey.
\newblock \emph{arXiv preprint arXiv:2402.06196}.

\bibitem[{Narvekar et~al.(2020)Narvekar, Peng, Leonetti, Sinapov, Taylor, and Stone}]{narvekar2020curriculum}
Sanmit Narvekar, Bei Peng, Matteo Leonetti, Jivko Sinapov, Matthew~E Taylor, and Peter Stone. 2020.
\newblock Curriculum learning for reinforcement learning domains: A framework and survey.
\newblock \emph{Journal of Machine Learning Research}, 21(181):1--50.

\bibitem[{Narvekar and Stone(2018)}]{narvekar2018learning}
Sanmit Narvekar and Peter Stone. 2018.
\newblock Learning curriculum policies for reinforcement learning.
\newblock \emph{arXiv preprint arXiv:1812.00285}.

\bibitem[{Qu et~al.(2025)Qu, Yang, Setlur, Tunstall, Beeching, Salakhutdinov, and Kumar}]{qu2025optimizing}
Yuxiao Qu, Matthew~YR Yang, Amrith Setlur, Lewis Tunstall, Edward~Emanuel Beeching, Ruslan Salakhutdinov, and Aviral Kumar. 2025.
\newblock Optimizing test-time compute via meta reinforcement fine-tuning.
\newblock \emph{arXiv preprint arXiv:2503.07572}.

\bibitem[{Ren et~al.(2024)Ren, Chen, Nguyen, Cui, Huang, and Yin}]{ren2024explicit}
Xuhui Ren, Tong Chen, Quoc Viet~Hung Nguyen, Lizhen Cui, Zi~Huang, and Hongzhi Yin. 2024.
\newblock Explicit knowledge graph reasoning for conversational recommendation.
\newblock \emph{ACM Transactions on Intelligent Systems and Technology}, 15(4):1--21.

\bibitem[{Rouditchenko et~al.(2025)Rouditchenko, Bhati, Araujo, Thomas, Kuehne, Feris, and Glass}]{rouditchenko2025omni}
Andrew Rouditchenko, Saurabhchand Bhati, Edson Araujo, Samuel Thomas, Hilde Kuehne, Rogerio Feris, and James Glass. 2025.
\newblock Omni-r1: Do you really need audio to fine-tune your audio llm?
\newblock \emph{arXiv preprint arXiv:2505.09439}.

\bibitem[{Sarwar et~al.(2001)Sarwar, Karypis, Konstan, and Riedl}]{sarwar2001item}
Badrul Sarwar, George Karypis, Joseph Konstan, and John Riedl. 2001.
\newblock Item-based collaborative filtering recommendation algorithms.
\newblock In \emph{Proceedings of the 10th international conference on World Wide Web}, pages 285--295.

\bibitem[{Schulman et~al.(2017)Schulman, Wolski, Dhariwal, Radford, and Klimov}]{schulman2017proximal}
John Schulman, Filip Wolski, Prafulla Dhariwal, Alec Radford, and Oleg Klimov. 2017.
\newblock Proximal policy optimization algorithms.
\newblock \emph{arXiv preprint arXiv:1707.06347}.

\bibitem[{Shao et~al.(2024)Shao, Wang, Zhu, Xu, Song, Bi, Zhang, Zhang, Li, Wu et~al.}]{shao2024deepseekmath}
Zhihong Shao, Peiyi Wang, Qihao Zhu, Runxin Xu, Junxiao Song, Xiao Bi, Haowei Zhang, Mingchuan Zhang, YK~Li, Yang Wu, and 1 others. 2024.
\newblock Deepseekmath: Pushing the limits of mathematical reasoning in open language models.
\newblock \emph{arXiv preprint arXiv:2402.03300}.

\bibitem[{Shi et~al.(2024)Shi, Deng, Luo, Xia, Bao, Ye, Du, Pan, and Li}]{shi2024llm}
Guangsi Shi, Xiaofeng Deng, Linhao Luo, Lijuan Xia, Lei Bao, Bei Ye, Fei Du, Shirui Pan, and Yuxiao Li. 2024.
\newblock Llm-powered explanations: Unraveling recommendations through subgraph reasoning.
\newblock \emph{arXiv preprint arXiv:2406.15859}.

\bibitem[{Team et~al.(2025)Team, Du, Gao, Xing, Jiang, Chen, Li, Xiao, Du, Liao et~al.}]{team2025kimi}
Kimi Team, Angang Du, Bofei Gao, Bowei Xing, Changjiu Jiang, Cheng Chen, Cheng Li, Chenjun Xiao, Chenzhuang Du, Chonghua Liao, and 1 others. 2025.
\newblock Kimi k1. 5: Scaling reinforcement learning with llms.
\newblock \emph{arXiv preprint arXiv:2501.12599}.

\bibitem[{Team(2024)}]{team2024qwen2}
Qwen Team. 2024.
\newblock Qwen2 technical report.
\newblock \emph{arXiv preprint arXiv:2407.10671}.

\bibitem[{Tsai et~al.(2024)Tsai, Kraft, Jin, Cai, Hosseini, Xu, Zhang, Hong, Chi, and Yi}]{tsai2024leveraging}
Alicia~Y Tsai, Adam Kraft, Long Jin, Chenwei Cai, Anahita Hosseini, Taibai Xu, Zemin Zhang, Lichan Hong, Ed~H Chi, and Xinyang Yi. 2024.
\newblock Leveraging llm reasoning enhances personalized recommender systems.
\newblock \emph{arXiv preprint arXiv:2408.00802}.

\bibitem[{Tu et~al.(2025)Tu, Feng, Chen, Liu, Tang, and Xie}]{tu2025vilbench}
Haoqin Tu, Weitao Feng, Hardy Chen, Hui Liu, Xianfeng Tang, and Cihang Xie. 2025.
\newblock Vilbench: A suite for vision-language process reward modeling.
\newblock \emph{arXiv preprint arXiv:2503.20271}.

\bibitem[{Wang et~al.(2025{\natexlab{a}})Wang, Karatzoglou, Arapakis, and Jose}]{wang2025large}
Jie Wang, Alexandros Karatzoglou, Ioannis Arapakis, and Joemon~M Jose. 2025{\natexlab{a}}.
\newblock Large language model driven policy exploration for recommender systems.
\newblock In \emph{Proceedings of the Eighteenth ACM International Conference on Web Search and Data Mining}, pages 107--116.

\bibitem[{Wang et~al.(2024{\natexlab{a}})Wang, Fang, Wan, Wen, Zhu, Liu, Gong, Song, Chen, Ni et~al.}]{wang2024openr}
Jun Wang, Meng Fang, Ziyu Wan, Muning Wen, Jiachen Zhu, Anjie Liu, Ziqin Gong, Yan Song, Lei Chen, Lionel~M Ni, and 1 others. 2024{\natexlab{a}}.
\newblock Openr: An open source framework for advanced reasoning with large language models.
\newblock \emph{arXiv preprint arXiv:2410.09671}.

\bibitem[{Wang et~al.(2025{\natexlab{b}})Wang, Wang, Chen, Hu, Girdhar, Wang, Gupta, Devella, Guo, Huang et~al.}]{wang2025adaptjobrec}
Qixin Wang, Dawei Wang, Kun Chen, Yaowei Hu, Puneet Girdhar, Ruoteng Wang, Aadesh Gupta, Chaitanya Devella, Wenlai Guo, Shangwen Huang, and 1 others. 2025{\natexlab{b}}.
\newblock Adaptjobrec: Enhancing conversational career recommendation through an llm-powered agentic system.
\newblock \emph{arXiv preprint arXiv:2508.13423}.

\bibitem[{Wang et~al.(2025{\natexlab{c}})Wang, Fan, Feng, Lin, Ma, Wang, and Yin}]{wang2025knowledge}
Shijie Wang, Wenqi Fan, Yue Feng, Shanru Lin, Xinyu Ma, Shuaiqiang Wang, and Dawei Yin. 2025{\natexlab{c}}.
\newblock Knowledge graph retrieval-augmented generation for llm-based recommendation.
\newblock \emph{arXiv preprint arXiv:2501.02226}.

\bibitem[{Wang et~al.(2024{\natexlab{b}})Wang, Zhang, Zhang, Hu, Li, Zhang, Li, Wu, Wang, and Hovy}]{wang2024reinforcement}
Shuhe Wang, Shengyu Zhang, Jie Zhang, Runyi Hu, Xiaoya Li, Tianwei Zhang, Jiwei Li, Fei Wu, Guoyin Wang, and Eduard Hovy. 2024{\natexlab{b}}.
\newblock Reinforcement learning enhanced llms: A survey.
\newblock \emph{arXiv preprint arXiv:2412.10400}.

\bibitem[{Wang et~al.(2025{\natexlab{d}})Wang, Du, Sun, Chua, Feng, Wang, and Zhang}]{wang2025re2llm}
Ziyan Wang, Yingpeng Du, Zhu Sun, Haoyan Chua, Kaidong Feng, Wenya Wang, and Jie Zhang. 2025{\natexlab{d}}.
\newblock Re2llm: reflective reinforcement large language model for session-based recommendation.
\newblock In \emph{Proceedings of the AAAI Conference on Artificial Intelligence}, volume~39, pages 12827--12835.

\bibitem[{Wei et~al.(2025)Wei, Yao, Liu, Zhang, Lu, Qiu, Yu, Xu, Zhang, Yin et~al.}]{wei2025webagent}
Zhepei Wei, Wenlin Yao, Yao Liu, Weizhi Zhang, Qin Lu, Liang Qiu, Changlong Yu, Puyang Xu, Chao Zhang, Bing Yin, and 1 others. 2025.
\newblock Webagent-r1: Training web agents via end-to-end multi-turn reinforcement learning.
\newblock \emph{arXiv preprint arXiv:2505.16421}.

\bibitem[{Williams(1992)}]{williams1992simple}
Ronald~J Williams. 1992.
\newblock Simple statistical gradient-following algorithms for connectionist reinforcement learning.
\newblock \emph{Machine learning}, 8(3):229--256.

\bibitem[{Xie et~al.(2025)Xie, Gao, Ren, Luo, Hong, Dai, Zhou, Qiu, Wu, and Luo}]{xie2025logic}
Tian Xie, Zitian Gao, Qingnan Ren, Haoming Luo, Yuqian Hong, Bryan Dai, Joey Zhou, Kai Qiu, Zhirong Wu, and Chong Luo. 2025.
\newblock Logic-rl: Unleashing llm reasoning with rule-based reinforcement learning.
\newblock \emph{arXiv preprint arXiv:2502.14768}.

\bibitem[{Yang et~al.(2024)Yang, Chen, and Fang}]{yang2024behavior}
Dayu Yang, Fumian Chen, and Hui Fang. 2024.
\newblock Behavior alignment: A new perspective of evaluating llm-based conversational recommendation systems.
\newblock In \emph{Proceedings of the 47th International ACM SIGIR Conference on Research and Development in Information Retrieval}, pages 2286--2290.

\bibitem[{Yu et~al.(2025)Yu, Zhang, Zhu, Yuan, Zuo, Yue, Dai, Fan, Liu, Liu et~al.}]{yu2025dapo}
Qiying Yu, Zheng Zhang, Ruofei Zhu, Yufeng Yuan, Xiaochen Zuo, Yu~Yue, Weinan Dai, Tiantian Fan, Gaohong Liu, Lingjun Liu, and 1 others. 2025.
\newblock Dapo: An open-source llm reinforcement learning system at scale.
\newblock \emph{arXiv preprint arXiv:2503.14476}.

\bibitem[{Zhang et~al.(2024)Zhang, Chen, Sheng, Wang, and Chua}]{zhang2024generative}
An~Zhang, Yuxin Chen, Leheng Sheng, Xiang Wang, and Tat-Seng Chua. 2024.
\newblock On generative agents in recommendation.
\newblock In \emph{Proceedings of the 47th international ACM SIGIR conference on research and development in Information Retrieval}, pages 1807--1817.

\bibitem[{Zhao et~al.(2024)Zhao, Fan, Li, Liu, Mei, Wang, Wen, Wang, Zhao, Tang et~al.}]{zhao2024recommender}
Zihuai Zhao, Wenqi Fan, Jiatong Li, Yunqing Liu, Xiaowei Mei, Yiqi Wang, Zhen Wen, Fei Wang, Xiangyu Zhao, Jiliang Tang, and 1 others. 2024.
\newblock Recommender systems in the era of large language models (llms).
\newblock \emph{IEEE Transactions on Knowledge and Data Engineering}, 36(11):6889--6907.

\bibitem[{Zhu et~al.(2025{\natexlab{a}})Zhu, Huang, and Sang}]{zhu2025llm}
Lixi Zhu, Xiaowen Huang, and Jitao Sang. 2025{\natexlab{a}}.
\newblock A llm-based controllable, scalable, human-involved user simulator framework for conversational recommender systems.
\newblock In \emph{Proceedings of the ACM on Web Conference 2025}, pages 4653--4661.

\bibitem[{Zhu et~al.(2025{\natexlab{b}})Zhu, Wan, Steck, Liang, Feng, Kallus, and Li}]{zhu2025collaborative}
Yaochen Zhu, Chao Wan, Harald Steck, Dawen Liang, Yesu Feng, Nathan Kallus, and Jundong Li. 2025{\natexlab{b}}.
\newblock Collaborative retrieval for large language model-based conversational recommender systems.
\newblock In \emph{Proceedings of the ACM on Web Conference 2025}, pages 3323--3334.

\bibitem[{Zou et~al.(2025)Zou, Yang, Chen, Hao, Chen, Huang, and Liang}]{zou2025traffic}
Xingchen Zou, Yuhao Yang, Zheng Chen, Xixuan Hao, Yiqi Chen, Chao Huang, and Yuxuan Liang. 2025.
\newblock Traffic-r1: Reinforced llms bring human-like reasoning to traffic signal control systems.
\newblock \emph{arXiv preprint arXiv:2508.02344}.

\end{thebibliography}

\appendix

\newpage
\appendix

\noindent\textbf{\LARGE APPENDIX}

\section{Dataset}
\label{app:dataset}
In this study, we conduct experiments based on RecBench+~\cite{huang2025towards}, which provides a clear difficulty hierarchy of user queries and covers diverse scenarios requiring complex reasoning, such as multi-hop reasoning and reflection. The queries are categorized into two main types: \textbf{Condition-based Query}, which encompasses hard constraints like directors or actors, and \textbf{User Profile-based Query}, which includes softer criteria such as user preferences or mood. Examples of these queries are presented in the table below:
\begin{table*}[h!]
    \centering
        \caption{Examples of RecBench+}
          \resizebox{1.0\textwidth}{!}{ 
    \begin{tabular}{c|p{10cm}|p{4cm}}
    
    \toprule
    Task & Query & Required Capability \\
    \midrule
    \multirow{4}[2]{*}{\makecell[c]{Explicit Condition\\Query}}
     & I’m really interested in classic films and would love to watch something that showcases \textbf{Charlie Chaplin}’s legendary comedic talent. Additionally, I’ve heard that \textbf{Roland Totheroh’s cinematography} adds an exceptional visual quality to movies. If you could point me in the direction of films that include both of these elements, I’d greatly appreciate it! & \textbf{Direct Reasoning:} Identifies specific attributes (e.g., director, cinematographer) and matches them directly. \\
     
    \midrule

    \multirow{4}[2]{*}{\makecell[c]{Implicit Condition\\Query}}
     & I recently \textbf{watched Clockers (1995)} and \textbf{Bamboozled (2000)}, and I was really impressed by the direction in both films. I'm eager to explore more works from the director, as I found their storytelling style and vision very engaging. If you could suggest other films \textbf{associated with this director}, that would be fantastic. & \textbf{Multi-hop Reasoning:} Infers the attribute from given information and then generate corresponding recommendations.\\
     
    \midrule

     \multirow{4}[2]{*}{\makecell[c]{Misinformed Condition\\Query}}
     & I recently watched \textbf{Lorenzo's Oil} and was really impressed by the cinematography done by \textcolor{red}{Mac Ahlberg}. I'm interested in finding more films that showcase his cinematographic style. I also remember seeing his work in \textbf{Beyond Rangoon}, so if there are any other movies he contributed to, I'd love to check them out! & \textbf{Reflection:} Detects and corrects misinformation (e.g., Mac Ahlberg did not work on these films) before generating recommendations. \\
     
    \midrule
     \multirow{3}[2]{*}{\makecell[c]{Interest-based\\Query}}
     & I'm fond of \textbf{romantic and dramatic} films from the \textbf{golden age of Hollywood} like \textquoteleft Roman Holiday' and \textquoteleft My Fair Lady'. Are there any other dramatic romances from that period you would recommend? & \textbf{Contextual Reasoning:} Leverages user interest context to suggest similar content. \\
     
     \midrule

      \multirow{2}[2]{*}{\makecell[c]{Demopraphics-based\\Query}}
     & I'm a \textbf{psychology professor} and I'm looking for movies that delve into \textbf{human emotions and relationships}. Have you got any? & \textbf{Domain-specific Reasoning:} Applies demographic details (e.g., occupation, age, gender) to recommend relevant content.\\
     
    \bottomrule
    \end{tabular}
    }
    \label{tab:query_example}
\end{table*}

\section{Baselines}
\label{app:baselines}
As traditional recommendation systems like GRU4Rec~\cite{hidasi2015session} and SASRec~\cite{kang2018self} cannot process natural language queries, in our experiments, we compare our method against three categories of approaches that can process natural language queries. Below is detailed information about these methods:
\begin{itemize}[leftmargin=0pt,itemsep=0pt, parsep=0pt, topsep=0pt, partopsep=0pt]
    \item \textbf{LLM Backbone:} This category employs pretrained LLM backbones directly for recommendation tasks. 
    \begin{itemize}
        \item \textbf{Qwen-2.5-3B-Instruct:} Qwen-2.5-3B-Instruct is a 3 billion parameter instruction-tuned LLM from the Qwen series, designed for enhanced performance in coding, mathematics, and instruction following tasks. 
        \item \textbf{Llama-3.2-3B-Instruct:} The Llama 3.2 collection of multilingual large language models (LLMs) is a collection of pretrained and instruction-tuned generative models in 1B and 3B sizes. The Llama 3.2 instruction-tuned text only models are optimized for multilingual dialogue use cases, including agentic retrieval and summarization tasks. 
        \item \textbf{DeepSeek-R1-Distill-Qwen-7B}: DeepSeek-R1-Distill-Qwen-7B is derived from Qwen-2.5 series, which are  finetuned with 800k samples curated with DeepSeek-R1.
        \item \textbf{GPT-4o:} GPT-4o is a multimodal LLM developed by OpenAI, capable of processing and generating text, images, and audio.
        \item \textbf{DeepSeek-R1~\cite{guo2025deepseek}:} DeepSeek-R1 is an open-source reasoning model developed by DeepSeek, trained using RL techniques to achieve state-of-the-art performance in reasoning tasks.
    \end{itemize}
    \item \textbf{LLM-based Conversational Recommender Systems (CRS):} These methods are LLM-based CRS designed for recommendation tasks. 
    \begin{itemize}
        \item \textbf{TallRec~\cite{bao2023tallrec}:} Utilizing LoRA technique to fine-tune LLMs on recommendation dataset.
        \item \textbf{InteRecAgent~\cite{huang2025recommender}:} InteRecAgent is a framework that combines LLMs with traditional recommender systems, enabling interactive and conversational recommendations by leveraging the strengths of both paradigms. It employs LLMs as the "brain" and uses recommender models as tools, incorporating components like memory, task planning, and reflection to transform traditional systems into interactive ones with natural language interfaces.
        \item \textbf{CRAG~\cite{zhu2025collaborative}:} CRAG is a conversational recommender system that combines LLMs with collaborative filtering techniques, providing context-aware and personalized recommendations.

    \end{itemize}
    \item \textbf{RFT-based Methods:} RLVR-based Methods include popular RL algorithms with verifiable rewards, focusing on improving LLM reasoning capabilities without extensive human annotation.
    \begin{itemize}
        \item \textbf{GRPO~\cite{shao2024deepseekmath}:} GRPO is an algorithm for training LLMs with RL. For each question, GRPO randomly sample multiple answers, and the advantage of an answer is defined as the normalized reward, thereby getting rid of the value estimation network.
        \item \textbf{REINFORCE++~\cite{hu2025reinforce++}:} REINFORCE++ is an enhanced variant of the classical REINFORCE~\cite{williams1992simple} algorithm, incorporating optimization techniques from Proximal Policy Optimization (PPO)~\cite{schulman2017proximal} while eliminating the need for a critic network.
        \item \textbf{RLOO~\cite{ahmadian2024back}:} RLOO (REINFORCE Leave One-Out) is a RL method that leaves out one sample for evaluation, ensuring robust alignment with human preferences.
        
    \end{itemize}
\end{itemize}

\section{Prompt}
\label{app:prompt}
We use below prompt template to guide the LLM in generating responses.
\begin{cmt*}{Prompt Template}{}
    Given the user's query, select one \textcolor{blue}{\{movie\textbackslash book\}} that best matches the query from candidates. You should think step-by-step and explain why you choose this \textcolor{blue}{\{movie\textbackslash book\}} and concisely why you didn't choose the others. The final answer should be in \textbackslash\textbackslash boxed\{\}.\\
    Query: \textcolor{blue}{\{query\}}\\
    Candidates: \textcolor{blue}{\{candidate list\}}
\end{cmt*}

\section{Implementation Details}
\subsection{RQ1}
\label{app:rq1}
Experiments were conducted on 2 H20 GPUs (96GB). The prompt template for the rollout process is detailed in \textbf{Appendix~\ref{app:prompt}}. For our method and all RFT-trained baselines, we use a learning rate of 5e-6, a group size of 5, and set the maximum response length to 768. Training is conducted for up to 15 epochs with early stopping (patience = 1). In our method, the hyperparameter $w_{penalty}$ in RAEE is set to 0.3. For the dual-graph-enhanced reward module, the weights $w_1$ and $w_2$ are set to equal values of 0.01 on the movie and book dataset. The difficulty threshold $\tau$ used in the online curriculum scheduler is set to 0.1.  The implementation utilized the PyTorch 2.6.0, Verl 0.3.1, VLLM 0.8.5, and Ray 2.46.0. The training-related hyperparameters are specified as follows: a batch size of 256 for policy updates, a KL loss coefficient of 0.01, a rollout temperature of 1.0, and a clipping ratio $\varepsilon$ of 0.2. For the movie and book domains, we sample 10,000 queries each as the training set and 12,000 queries each as the test set. Training is conducted separately on these two domains.

\subsection{RQ2}
We select Condition-based Queries that are not present in the training set. For each query, the candidates are constructed as follows: one positive item, three hard negatives (items that satisfy the conditions in the query but are not present in the user's interaction history), and sixteen simple negatives (randomly sampled items).

For the "without history" setting, we use the prompt from Appendix~\ref{app:prompt} to allow the trained \ourmodel model (with Qwen-2.5-3B-Instruct as the backbone) to make predictions. In the "with history" setting, we build upon the previous prompt and introduce the user's \{watching/reading\} history before the candidates. This history consists of 10 randomly sampled items from the user interaction history.

\subsection{RQ3}
\label{app:rq3}
\textbf{Cross-Domain Setting:}
For the cross-domain evaluation, we directly apply the model trained on the Movie domain to make predictions in the Book domain, and vice versa. As a baseline, we include the original base model without any task-specific training.

\noindent
\textbf{Cross-Task Setting:}
To assess generalization across tasks, we apply \ourmodel—trained on our task—to the sequential recommendation task. In this setup, the model is given a user's interaction history and asked to predict the next item. We compare performance against the base model and standard recommendation baselines, including GRU4Rec~\cite{hidasi2015session} and SASRec~\cite{kang2018self}. The maximum length of the interaction history is set to 10, and all models are required to select the most likely item from a pool of 20 candidates.

\subsection{RQ4}
\label{app:rq4}
\textbf{Supervised Fine-Tuning (SFT) Details:}
We fine-tuned the Qwen-2.5-3B-Instruct model using full parameter training, with the input provided by Prompt~\ref{app:prompt} and the corresponding target items as the output. Training was conducted on an H20 GPU (96GB memory) using a learning rate of 1e-5. We applied early stopping with a patience of 1 to prevent overfitting.

\textbf{Evaluation Details:}
We evaluated the Qwen-2.5-3B-Instruct, the fine-tuned SFT model, and \ourmodel across a series of benchmarks. All models were tested under identical inference settings to ensure fair comparison. Below is the information of the used benchmarks:
\begin{itemize}[leftmargin=*, itemsep=0pt, parsep=0pt, topsep=0pt, partopsep=0pt]
    \item \textbf{DROP:} The DROP~\footnote{https://modelscope.cn/datasets/AI-ModelScope/DROP/summary} (Discrete Reasoning Over Paragraphs) benchmark is designed to evaluate the reading comprehension and reasoning capabilities of AI models. It includes a variety of tasks that require models to read passages and answer questions based on the content. We evaluate models with zero-shot setting and use accuracy as evaluation metric.
    \item \textbf{IFEval:} IFEval~\footnote{https://modelscope.cn/datasets/opencompass/ifeval/summary} is a benchmark for evaluating instruction-following language models, focusing on their ability to understand and respond to various prompts. It includes a diverse set of tasks and metrics to assess model performance comprehensively. We evaluate models with zero-shot setting and use accuracy as evaluation metric.
    \item \textbf{ARC:} ARC (AI2 Reasoning Challenge)~\footnote{https://modelscope.cn/datasets/modelscope/ai\_arc/summary} benchmark is designed to evaluate the reasoning capabilities of AI models through multiple-choice questions derived from science exams. It includes two subsets: ARC-Easy and ARC-Challenge, which vary in difficulty. We evaluate models with zero-shot setting and use accuracy as evaluation metric.
    \item \textbf{GPQA:} GPQA~\footnote{https://huggingface.co/datasets/Idavidrein/gpqa} is a multiple-choice, Q\&A dataset of very hard questions written and validated by experts in biology, physics, and chemistry. When attempting questions out of their own domain (e.g., a physicist answers a chemistry question), these experts get only 34\% accuracy, despite spending >30 minutes with full access to Google. We evaluate models with a 5-shot setting and use pass@1 as an evaluation metric.

\end{itemize}

\section{Parameter Analysis}

\begin{figure}[h]
    \centering
    \includegraphics[width=0.75\linewidth]{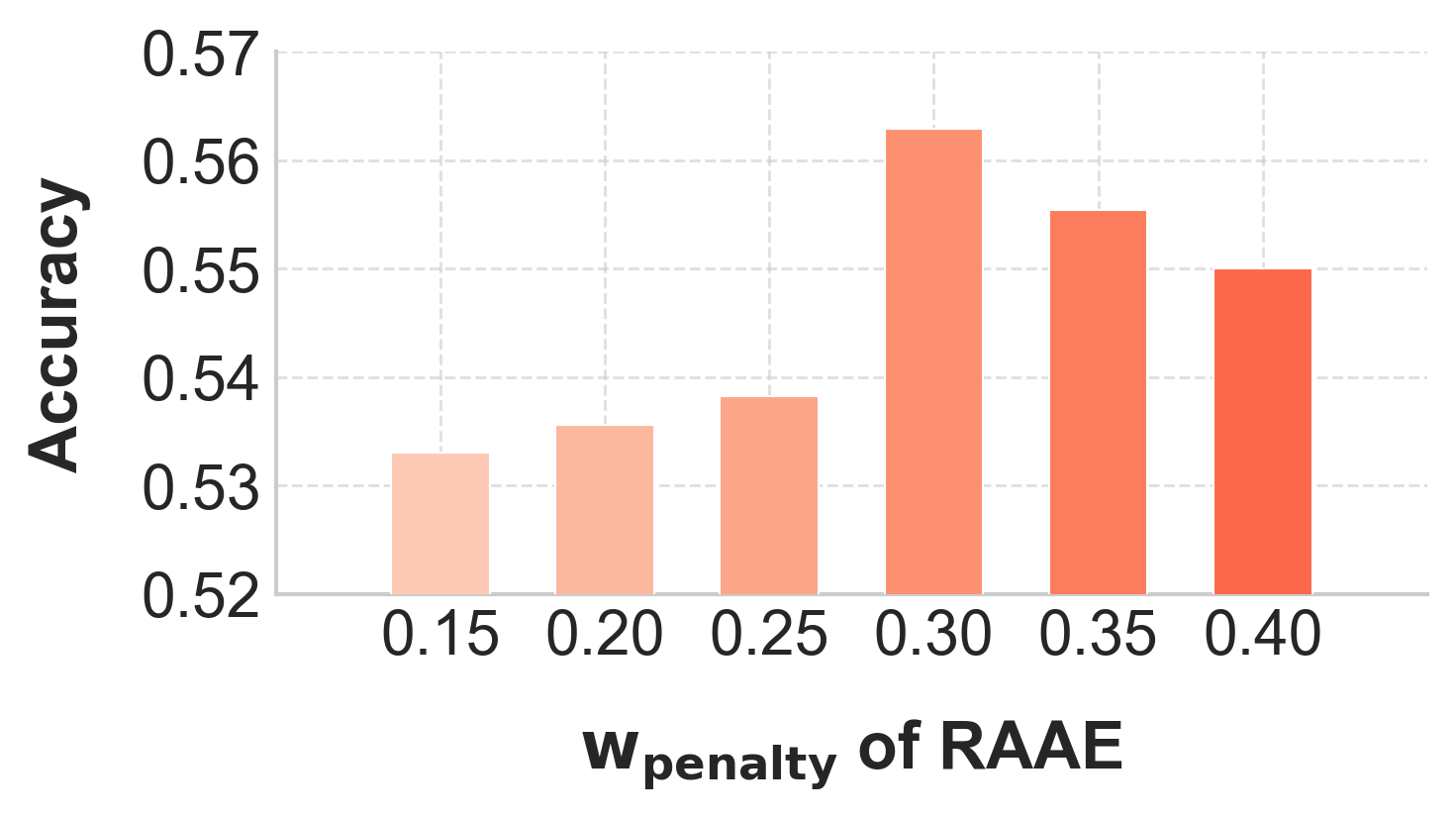}
    \caption{Effect of $w_{\text{penalty}}$ of RAAE
    }
    \label{fig:w_penalty}
\end{figure}
In this section, we further explored the influence of the penalty parameter $w_{\text{penalty}}$ for incorrect reasoning steps in RAEE on model performance. As shown in Figure~\ref{fig:w_penalty}, accuracy improves as $w_{\text{penalty}}$ increases, peaking at 0.30, after which it gradually drops. This suggests that moderate penalization helps suppress incorrect reasoning, while overly strong penalties may harm final performance. Consequently, we set $w_{\text{penalty}}=0.3$ as the default in this paper to achieve the best performance.

\begin{figure*}[t]
    \centering
\begin{cmt*}{Case for \ourmodel Reasoning }{}

\textbf{Query:} I recently watched \underline{Pan's} \underline{Labyrinth} \underline{(El Laberinto del fauno)} \underline{(2006)} and \underline{Four Rooms (1995)} and was really captivated by the cinematography. The visual storytelling in both films left a strong impression on me, and I’m eager to explore more movies that share this distinctive visual style. If you could suggest films that were \underline{shot by the same cinematographer}, that would be great!
\textbf{Candidates:} Stuart Little 2 (2002);Fabulous Baker Boys, The (1989);Hellboy (2004);Last Supper, The (1995);Twin Dragons (Shuang long hui) (1992);Go (1999);Escape from New York (1981);Hanging Garden, The (1997);They Made Me a Criminal (1939);King Kong (1933);Terms of Endearment (1983);Right Stuff, The (1983);\textbf{\textcolor{lightgreen}{The Long Kiss Goodnight (1996)}};They Bite (1996);Home Alone 2: Lost in New York (1992);Tora! Tora! Tora! (1970);Pan's Labyrinth (El Laberinto del fauno) (2006);Devil Girl From Mars (1954);Van, The (1996);Iron Eagle IV (1995)
\vspace{10pt}
\\

\textbf{\ourmodel: }I understand the user's preference for visually striking films and the specific cinematography style mentioned. To narrow down the options, I'll first identify which cinematographers are known for utilizing striking visual storytelling and cinematography. One strong candidate here could be \underline{Guillermo Navarro}, who has directed both films that the user liked. He is known for his visually distinctive works.

Given the candidates:
\begin{itemize}
    \item Guillermo Navarro is associated with films like \textbf{\textcolor{lightgreen}{The Long Kiss Goodnight (1996)}}.
    \item \textbf{\textcolor{lightred}{Hellboy (2004)}} also have Guillermo Navarro as the director, but its style is not very similar to the two movies mentioned by the user, so it may not be the best choice.
\end{itemize}

\textbf{\textcolor{lightgreen}{The Long Kiss Goodnight (1996)}} is the movie I believe best portrays the distinctive visual style mentioned by the user and credits the same cinematography team as both films the user was captivated by. Other options were not as relevant to the context and criteria specified.

\end{cmt*}
\end{figure*}

\section{Case Study}

To intuitively demonstrate \ourmodel’s reasoning capabilities on \ourtask, we present a representative case study below. The user query seeks films shot by the same cinematographer as Pan’s Labyrinth (2006) and Four Rooms (1995), emphasizing their distinctive visual storytelling. \ourmodel delivers a precise recommendation through the following step-by-step reasoning process, showcasing its ability to handle complex, context-driven queries effectively.

\begin{itemize}[leftmargin=*,itemsep=0pt, parsep=0pt, topsep=0pt, partopsep=0pt]
    \item 
\textbf{Multi-hop Reasoning:} \ourmodel correctly identifies \textbf{Guillermo Navarro} as the cinematographer for \textit{Pan’s Labyrinth} and links his work to \textit{The Long Kiss Goodnight (1996)} and \textit{Hellboy (2004)} from the candidate list.

\item \textbf{Contextual Evaluation:} Beyond merely matching cinematographers, \ourmodel evaluates stylistic alignment, dismissing \textit{Hellboy (2004)}—another Navarro film—due to its divergence from the visual style of the user’s preferred films. This reflects a nuanced understanding of user intent and context. 
\end{itemize}

\section{Future Work}
\label{app:future}

While ReRec establishes a robust framework for enhancing reasoning capabilities in LLM-based recommendation assistants, several promising directions remain open for future exploration.

\textbf{Multi-turn Conversational Recommendation with Reasoning.}
One compelling future direction is extending ReRec to multi-turn conversational settings, where user intents evolve dynamically through dialogue. In such scenarios, the task becomes significantly more complex: the model must track contextual dependencies, resolve coreference, and integrate intermediate reasoning steps across utterances. Designing effective reward models under this setting is non-trivial. Unlike single-turn queries, multi-turn conversations require the model to maintain dialogue coherence and long-term goal alignment. Future work could explore hierarchical reward structures that combine utterance-level relevance with cumulative dialogue success.

\textbf{Reasoning with Tools for Recommendation.}
Another important direction is augmenting ReRec with external tools to enable tool-augmented reasoning. While LLMs possess broad linguistic and general reasoning capabilities, real-world recommendation often requires interfacing with external systems—such as retrieval engines, structured databases, or collaborative filtering models—to access up-to-date or user-specific information. To this end, future research could explore LLM-agent architectures for recommendation, where the model acts as a planner that issues sub-tasks to external tools and integrates the responses via step-by-step reasoning. 


\end{document}